 \documentclass{aastex6}

\fullcollaborationName{The Friends of AASTeX Collaboration}

\shorttitle{}
\shortauthors{Katsuda et al.}

\begin{document}


\title{Intermediate-Mass-Elements in Young Supernova Remnants Reveal Neutron Star Kicks by Asymmetric Explosions}


\author{Satoru Katsuda\altaffilmark{1,2}, Mikio Morii\altaffilmark{3}, 
Hans-Thomas Janka\altaffilmark{4}, Annop Wongwathanarat\altaffilmark{4,5},\\
Ko Nakamura\altaffilmark{6}, Kei Kotake\altaffilmark{6}, 
Koji Mori\altaffilmark{7}, Ewald M\"uller\altaffilmark{4}, 
Tomoya Takiwaki\altaffilmark{8}, Masaomi Tanaka\altaffilmark{8},\\ 
Nozomu Tominaga\altaffilmark{9,10}, Hiroshi Tsunemi\altaffilmark{11}
}

\altaffiltext{1}{Graduate School of Science and Engineering, Saitama University, 255 Shimo-Ohkubo, Sakura, Saitama 338-8570, Japan; katsuda@phy.saitama-u.ac.jp}

\altaffiltext{2}{Department of Physics, Faculty of Science \& Engineering, Chuo University, 1-13-27 Kasuga, Bunkyo, Tokyo 112-8551, Japan}

\altaffiltext{3}{The Institute of Statistical Mathematics, 10-3 Midori-cho, Tachikawa, Tokyo 190-8562, Japan; morii@ism.ac.jp}

\altaffiltext{4}{Max-Planck-Institut f\"ur Astrophysik, Karl-Schwarzschild-Str. 1, D-85748 Garching, Germany; thj@mpa-garching.mpg.de}

\altaffiltext{5}{RIKEN, Astrophysical Big Bang Laboratory, 2-1 Hirosawa, Wako, 351-0198, Saitama, Japan}

\altaffiltext{6}{Department of Applied Physics, Fukuoka University, Jonan, Nanakuma, Fukuoka 814-0180, Japan}

\altaffiltext{7}{Department of Applied Physics and Electronic Engineering, University of Miyazaki, 1-1, Gakuen Kibanadai-nishi, Miyazaki 889-2192, Japan}

\altaffiltext{8}{National Astronomical Observatory of Japan, Osawa, Mitaka, Tokyo 181-8588, Japan}

\altaffiltext{9}{Department of Physics, Faculty of Science and Engineering, Konan University, 8-9-1 Okamoto, Kobe, Hyogo 658-8501, Japan}

\altaffiltext{10}{Kavli Institute for the Physics and Mathematics of the Universe (WPI), The University of Tokyo, 5-1-5 Kashiwanoha, Kashiwa, Chiba 277-8583, Japan}

\altaffiltext{11}{Department of Earth and Space Science, Graduate School of Science, Osaka University, 1-1 Machikaneyama, Toyonaka, Osaka 560-0043, Japan}

\begin{abstract}

The birth properties of neutron stars yield important information on the still debated physical processes that trigger the explosion and on intrinsic neutron-star physics.  These properties include the high space velocities of young neutron stars with average values of several 100\,km\,s$^{-1}$, whose underlying ``kick" mechanism is not finally clarified.  There are two competing possibilities that could accelerate NSs during their birth: anisotropic ejection of either stellar debris or neutrinos.  We here present new evidence from X-ray measurements that chemical elements between silicon and calcium in six young gaseous supernova remnants are preferentially expelled opposite to the direction of neutron star motion.  There is no correlation between the kick velocities and magnetic field strengths of these neutron stars.  Our results support a hydrodynamic origin of neutron-star kicks connected to asymmetric explosive mass ejection, and they conflict with neutron-star acceleration scenarios that invoke anisotropic neutrino emission caused by particle and nuclear physics in combination with very strong neutron-star magnetic fields.  

\end{abstract}

\keywords{ISM: supernova remnants --- stars: neutron --- methods: data analysis --- techniques: imaging spectroscopy --- X-rays: general}

\section{Introduction} \label{sec:intro}

Neutron stars (NSs) are the most compact stars in the Universe.  They are formed in violent explosions terminating the lives of massive stars, i.e., core-collapse supernovae (SNe).  NSs are exotic and fascinating objects because of many peculiar properties such as extremely high densities, strong magnetic fields, and rapid rotation.  In addition, NSs achieve high velocities of several 100 km\,s$^{-1}$ on average at birth \citep{2005MNRAS.360..974H}.  The origin of their high velocities has been a long-standing mystery in astrophysics \citep{2001ApJ...549.1111L}.   There are two competing mechanisms that could accelerate NSs during their birth: anisotropic ejection of the stellar debris \citep[``hydrodynamic kicks":][]{1994A&A...290..496J,1996PhRvL..76..352B} or asymmetric-neutrino emission \citep[``neutrino-induced kicks":][]{1987IAUS..125..255W,1993AAT....3..287B,2005ApJ...632..531S}.  The origin of their high velocities has been a long-standing mystery in astrophysics.  It cannot be explained by the disruption of close binaries as a consequence of the second SN explosion \citep{1961BAN....15..265B}, but it requires intrinsic kicks during the SN blast itself \citep[see][and references therein]{2001ApJ...549.1111L}.

An asymmetry of 3\% of the total neutrino emission, whose energy is typically several 10$^{53}$ erg, is sufficient to accelerate a NS of 1.5 solar masses (M$_\odot$) to a velocity of 1000 km\,s$^{-1}$.  But even such a small emission asymmetry is hard to generate in the strong gravitational field of a NS.  Extremely strong dipole magnetic fields ($\gtrsim$10$^{16}$\,G) and/or special assumptions for the neutrino and nuclear physics in the NS interior must be invoked for its creation, e.g., modifications of the neutrino interactions in strongly magnetized hadronic, hyperonic or quark matter, keV sterile neutrinos or the neutrino-bubble instability \citep[e.g.,][]{1987IAUS..125..255W,1993AAT....3..287B,2005ApJ...632..531S,2003PhRvD..68j3002F,2008A&A...489..281S,2012PhRvD..86l3003M}.  Such models either suggest no correlation between SN explosion asymmetries and NS kick velocities, or they predict strongest mass ejection in the direction of the NS motion \citep{2006ApJS..163..335F}.  

Hydrodynamic kicks, on the other hand, appear to be a natural consequence of the large-scale asphericities seen in most core-collapse SNe \citep[e.g.,][]{2000ApJS..127..141N,2006Natur.440..505L,2008Sci...319.1220M} and their gaseous remnants \citep[e.g.,][]{2011ApJ...732....3R,2014Natur.506..339G}.  In fact, recent two- and three-dimensional (2D, 3D) hydrodynamic simulations of the neutrino-driven mechanism, in which energy transfer by neutrinos powers the SN explosion, produce NS kicks up to more than 1000 km\,s$^{-1}$ \citep{2006A&A...457..963S,2013A&A...552A.126W,2010PhRvD..82j3016N,2012MNRAS.423.1805N}.  The NS velocities are directed opposite to the hemisphere of the stronger explosion, where the explosively nucleosynthesized elements from Si to Fe are preferentially expelled.  A clear correlation between the asymmetry of these innermost ejecta and the magnitude of the kick velocity is predicted \citep{2013A&A...552A.126W}.  Quantitative predictions, however, require a detailed understanding of the still disputed physical processes that power the explosion.  Therefore, observationally establishing a connection between SN asymmetries and NS kicks can help to constrain theoretical models for successful supernova explosions.

Supernova remnants (SNRs) in our own Galaxy offer a unique opportunity to test these predictions and to discriminate between the two competing NS kick mechanisms, because their proximity allows one to reveal the detailed morphology of the SN debris and to determine the precise position of the NS.  Recent detailed X-ray mapping of Galactic SNRs such as Cassiopeia~A (Cas~A hereafter) and G292.0+1.8 has revealed that the bulk of the total ejecta (dominated by oxygen) as well as the innermost ejecta of $^{44}$Ti travel roughly in the direction opposite to the apparent motion of the NS \citep{2014Natur.506..339G,2012ApJ...746..130H,2017ApJ...844...84H}.  These cases provide us with encouraging supports for the hydrodynamic-kick mechanism, whereas a systematic analysis of the ejecta component for bigger SNR samples is demanded to claim convincing evidence.  

Recently, \citet{2017ApJ...844...84H} systematically measured asymmetries of X-ray morphologies for 18 SNRs with {\it Chandra} and {\it ROSAT}.  They found that NSs are preferentially moving opposite to the bulk of the X-ray emission, supporting the hydrodynamic-kick scenario.  However, there remain several caveats in their analyses.  First, they measured the asymmetries in the 0.5--2.1\,keV band without separating the SN ejecta and the swept-up circumstellar medium (CSM) components, although they do discuss the relative contributions of these components based on past X-ray studies.  Second, their proper-motion measurements of NSs, which are essential to infer the explosion sites and thus affect both the NS kick direction and the SN asymmetry, are subject to systematic uncertainties that were not considered in their work.  For example, they derived a very strict constraint on the proper motion of the NS in RCW~103 to be $\sim$0.010$\pm0.003^{\prime\prime}$\,yr$^{-1}$ for a time baseline of 16\,yrs.  This uncertainty is an order of magnitude smaller than that of the other proper motion measurement for the NS in Puppis~A, with a time baseline of 10.5\,yrs \citep[0.071$\pm$0.034$^{\prime\prime}$\,yr$^{-1}$:][]{2012ApJ...755..141B}.  Large systematic uncertainties of a few 0.1$^{\prime\prime}$ have been also quoted on expansion measurements of SNRs \citep[e.g.,][]{2010ApJ...709.1387K}.  These considerations lead us to the conclusion that the work by \citet{2017ApJ...844...84H} is not yet conclusive.  Here we present results based on a more sophisticated analysis for six young SNRs.

\section{Target Selection and Data}

We selected six young core-collapse SNRs associated with NSs, namely Cas~A, G292.0+1.8, Puppis~A, Kes~73, RCW~103, and N49.  Only these samples pass the following criteria: (i) relatively young ($\lesssim$ several kyr), (ii) inter-mediate-mass (IME defined as Si, S, Ar, and Ca) ejecta are detected in X-rays, and (iii) neutron stars are detected within the gaseous remnant.  It should be noted that our analysis focused on lighter IMEs such as Si and S, as can be seen in the following figures and tables.  According to the theoretical models \citep{2013A&A...552A.126W}, Fe is a better tracer of explosive burning and is more sensitive to the NS kick than these lighter IMEs.  However, there are two problems with Fe.  For young SNRs like Cas~A, much of Fe may not be heated by reverse shocks, as Fe is thought to be located in the innermost part of SNRs.  This would make it difficult to measure the distribution of Fe.  Whereas most of Fe could be heated by reverse shocks in evolved SNRs, the plasmas are usually not hot enough to produce Fe K line emission.  In these cases, Fe abundances are measured by Fe L line emission which suffers from incomplete atomic data.  Therefore, we here focus on IMEs rather than Fe.

Table~\ref{tab:targets} summarizes basic properties of the selected six SNRs as well as their associated NSs.  The first three SNRs are well-studied cases, in which fast-moving stellar debris (mainly O-rich ejecta) is identified in the optical.   This allows one to precisely estimate the position of the SN explosion, so-called ``center of expansion (CoE)", by tracing back proper motions of the fast-moving debris \citep{2006ApJ...645..283F,2009ApJ...692.1489W,1988srim.conf...65W}.  The other remnants are known to host strongly-magnetized ($\gtrsim$10$^{14}$\,G) neutron stars, so-called magnetars, AXP 1E1841-045, 1E161348-5055, and SGR 0526-66, respectively.  We used archival data acquired with {\it Chandra} and {\it XMM-Newton}, as listed in Table~\ref{tab:data}.  We performed a standard data processing with the latest calibration database at the analysis phase. 

\section{Analysis}

To reveal whether or not NSs recoil to the expelled heavy elements, we need to measure (i) centers of mass (CoM) of heavy-element ejecta, (ii) explosion sites, and (iii) positions of NSs and their proper motions, if available.  As in Table~\ref{tab:targets}, preceding works have already measured not only the NS positions of all the six sources but also the explosion sites for Cas~A, Puppis~A, and G292.0+1.8.  Therefore, the CoM is a particularly important parameter that we must measure in this work.  Additionally we inferred the explosion sites for Kes~73, RCW~103, and N49, based on the geometric center of the X-ray boundaries (CoX).

\subsection{Measurement of the CoM of the IME-rich ejecta}

\subsubsection{Image Decomposition}

X-ray emission generally consists of several distinct spectral components, such as a swept-up CSM, SN ejecta, and power-law (PL).  Therefore, it is fundamentally important to separate these components spectroscopically.  To this end, we introduce an efficient technique to decompose raw data taken by X-ray CCD cameras onboard {\it Chandra} and {\it XMM-Newton} into maps of different spectral components.  The method reduces both machine time and human labor, potentially enabling one to reveal ejecta distributions for a number of SNRs.

The X-ray data taken by {\it Chandra} and {\it XMM-Newton}, in principle, allow one to generate an X-ray energy spectrum from every single pixel.  Our decomposition method is to fit these spectra with several spectral components by allowing only their relative contributions (i.e., normalizations) to vary freely.  In other words, we assume that the number of photons ($n_i$) at the $i$-th energy bin $E_i$ ($i = 1, ..., N$; we binned the 0.5--7 keV band into 40 steps logarithmically when performing the image decomposition) in each pixel can be modeled with a linear combination of spectral model functions $M_j$ ($E_i$) ($j = 1, ..., M$) by 
\begin{equation}
  n_i \simeq f(E_i; {\bf k} ) = \sum_{j = 1}^{M} k_j M_j(E_i).
\end{equation}
where ${\bf k} = (k_1, ..., k_M)^T$.  As we describe below, the spectral model functions are constructed through standard spectral analyses with conventional models \citep[e.g., the {\tt vpshock} model for the thermal emission:][]{2001ApJ...548..820B} in the XSPEC package \citep{1996ASPC..101...17A}.  We took the value of $n_i$ from narrow-band images after correcting for vignetting effects.  We assumed the same template spectral model throughout the remnant, which is equivalent to an assumption that there is no spatial variation in the spectral response except for the effective area.  This is subject to a systematic uncertainty of our image decomposition.  However, the spectral variation of the spectral response (i.e., energy resolution and energy scale) is within a few 10\% within the same CCD chip\footnote{http://cxc.harvard.edu/cal/Acis/detailed\_info.html}, and thus this assumption would not affect our final conclusion on the image decomposition.  We note that G292.0+1.8 and RCW~103 spread to multiple CCD chips.  In these cases, the energy resolution could vary by $\sim$20\%.  Nonetheless, given that our spectral binning is fairly large (1 bin = $\sim$120\,eV at Si He$\alpha$), it is highly unlikely that the difference in spectral response affects the result of our image decomposition.  

Based on the literature, we identified dominant X-ray spectral components for each SNR as follows: IME-rich ejecta, Fe-rich ejecta, O-rich ejecta, CSM, and power-law for Cas~A \citep{2012ApJ...746..130H}; IME-rich ejecta, O-rich ejecta, CSM, power-law for G292.0+1.8 \citep{2004ApJ...602L..33P}; IME-rich ejecta, O-rich ejecta, CSM with relatively strong absorption, and CSM with relatively weak absorption for Puppis A \citep{2008ApJ...676..378H,2008ApJ...678..297K,2010ApJ...714.1725K,2013ApJ...768..182K}; IME-rich ejecta and CSM for Kes~73, RCW~103, and N49 \citep{2014ApJ...781...41K,2015ApJ...810..113F,2012ApJ...748..117P}.  In principle, we can include additional components whose presence had not been reported.  For example, a power-law component might be present not only in Cas~A and G292.0+1.8 but also in the other SNRs.  However, adding such components could cause incorrect decompositions in some cases, e.g., a power-law component negatively impacts the other components unreasonably.  Therefore, we utilized only spectra whose presence had been established by preceeding X-ray studies.

\begin{figure}[ht!]
\gridline{\fig{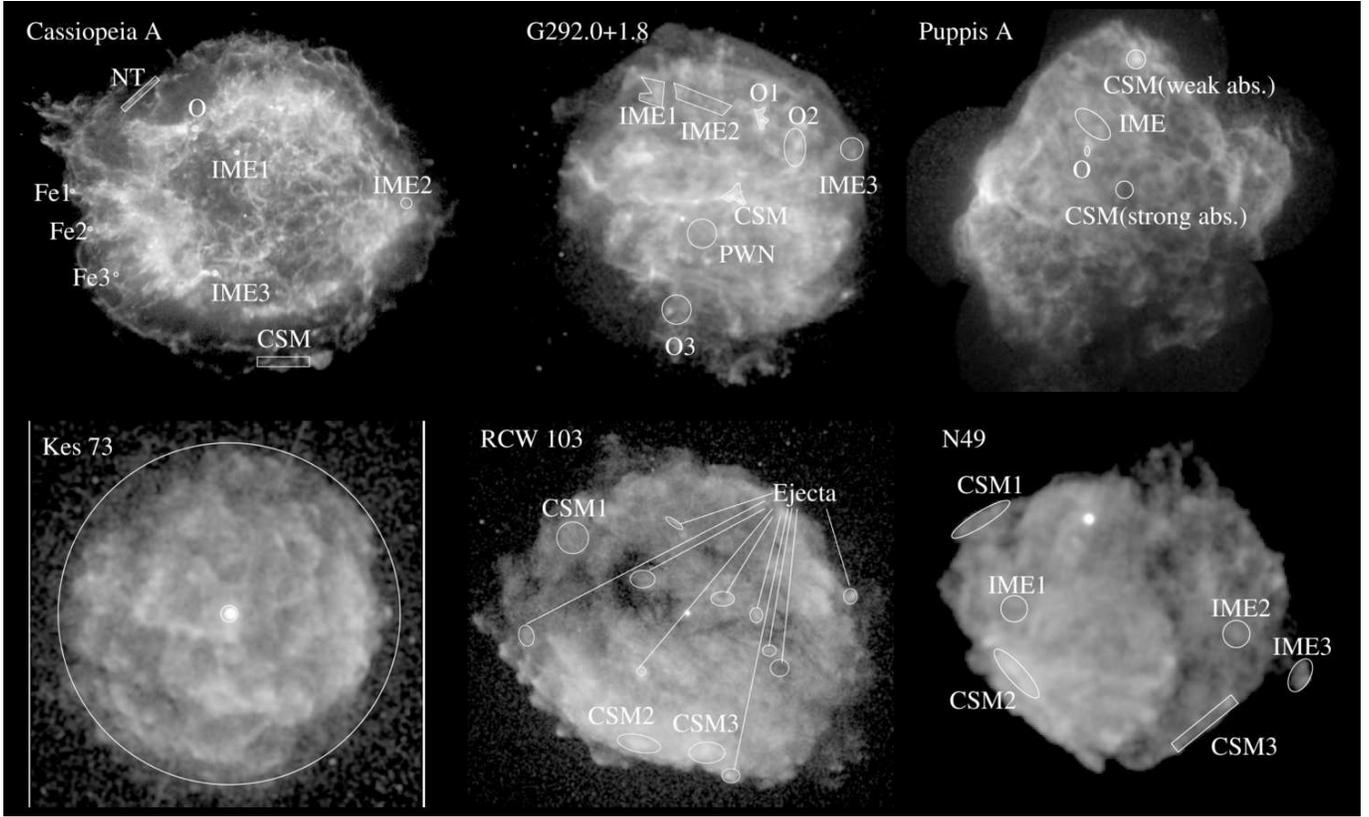}{1.0\textwidth}{}
          }
\caption{Spectral extraction regions to generate template spectra for our image decomposition.  We analyzed the whole remnant excluding the neutron star for Kes~73, whereas we picked up several small regions for the other five remnants.  See text for more details.
\label{fig:specregs}}
\end{figure}

\begin{figure}[ht!]
\gridline{\fig{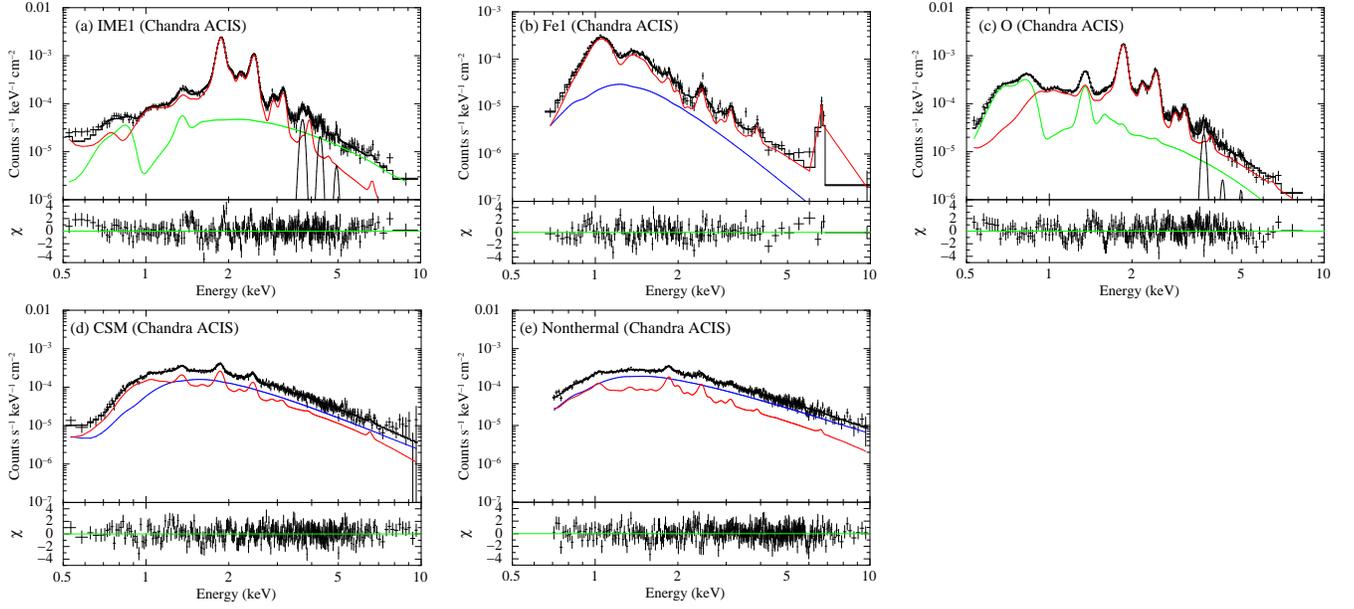}{1.0\textwidth}{}
          }
\caption{Five representative spectra in Cas~A, together with the best-fit model components and the residuals in the lower panels. The data and model components are shown in black crosses and solid lines, respectively. The individual components are as follows. (a) and (c) Red, green, and black lines, respectively, represent the IME-rich ejecta, the O-rich ejecta, and additional Gaussian components to represent pile-up effects for prominent Si He$\alpha$, Si Ly$\alpha$, and S He$\alpha$. (b) Red and blue represent the Fe-rich ejecta and power-law components, respectively. (d)-(e) Red and blue represent CSM and power-law components, respectively.
\label{fig:casa_spec}}
\end{figure}

\begin{figure}[ht!]
\gridline{\fig{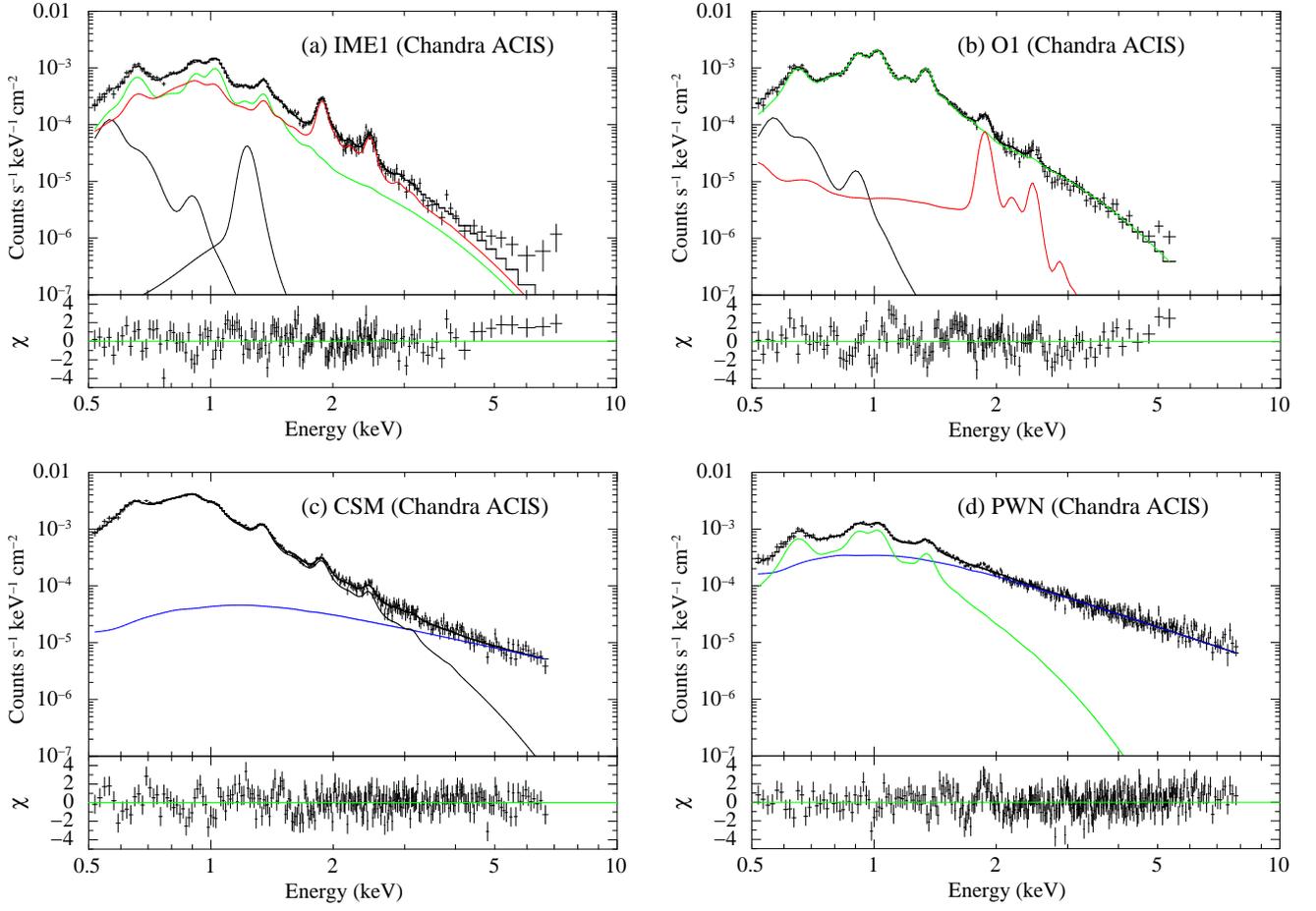}{1.0\textwidth}{}
          }
\caption{Same as Fig.~\ref{fig:casa_spec}, but for G292.0+1.8. (a)-(b) Red, green, and black lines represent IME-rich ejecta, O-rich ejecta, and CSM components, respectively. (c) Black and blue lines represent CSM and power-law components, respectively. (d) Blue and green lines represent power-law and O-rich ejecta components, respectively.
\label{fig:g292_spec}}
\end{figure}

\begin{figure}
\gridline{\fig{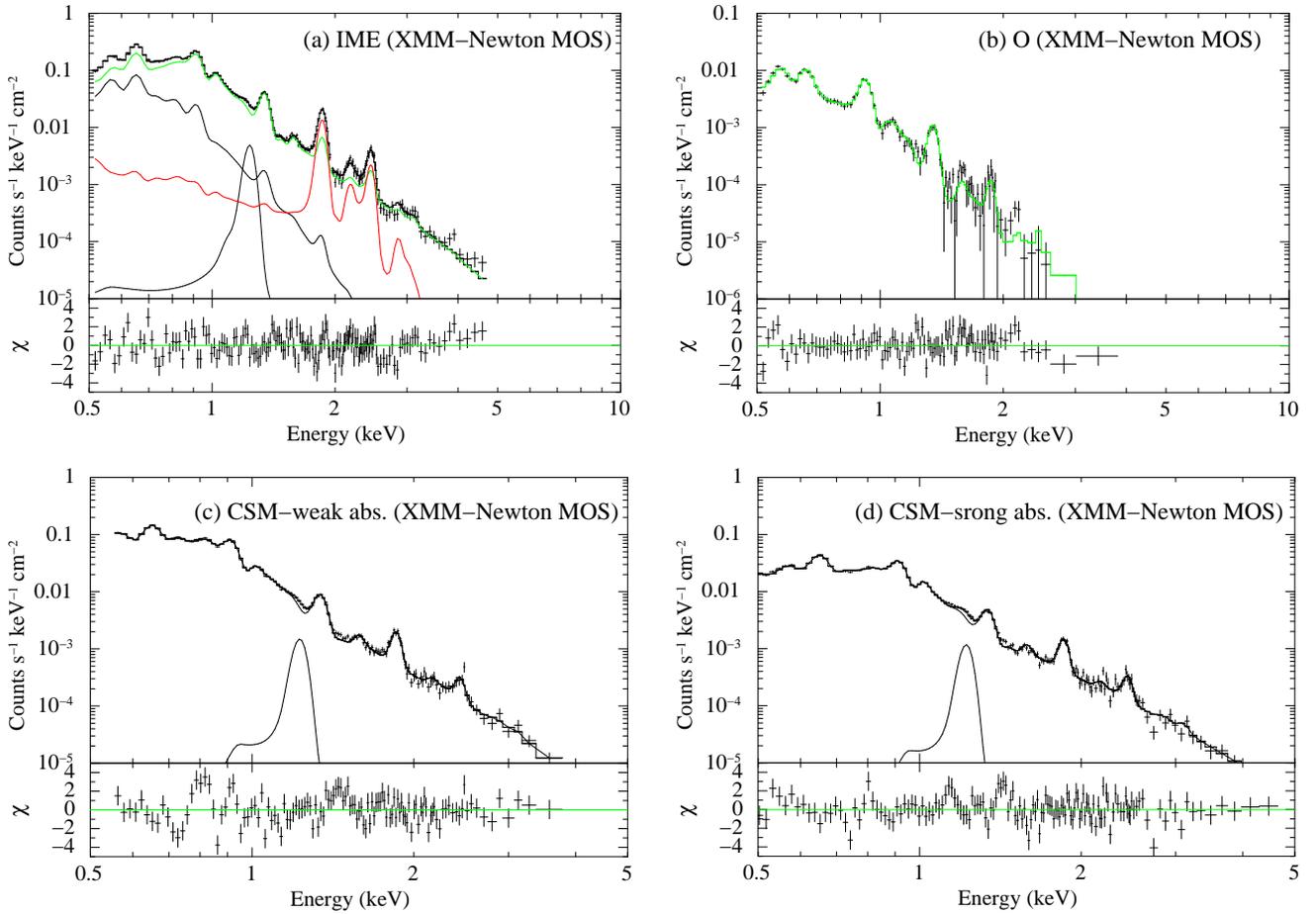}{1.0\textwidth}{}
          }
\caption{Same as Fig.~\ref{fig:casa_spec}, but for Puppis~A. (a) Red, green, and black lines represent IME-rich ejecta, O-rich ejecta, and CSM components, respectively. The black Gaussian at $\sim$1.2 keV has been added to reproduce missing line emission (presumably Fe L lines) in the model. (b) We extracted a pure O-rich ejecta spectrum by subtracting a local-background. (c) Same as (a) with a weakly absorbed CSM component. (d) Same as (a) with a strongly absorbed CSM component.
\label{fig:pupa_spec}}
\end{figure}

\begin{figure}[ht!]
\gridline{\fig{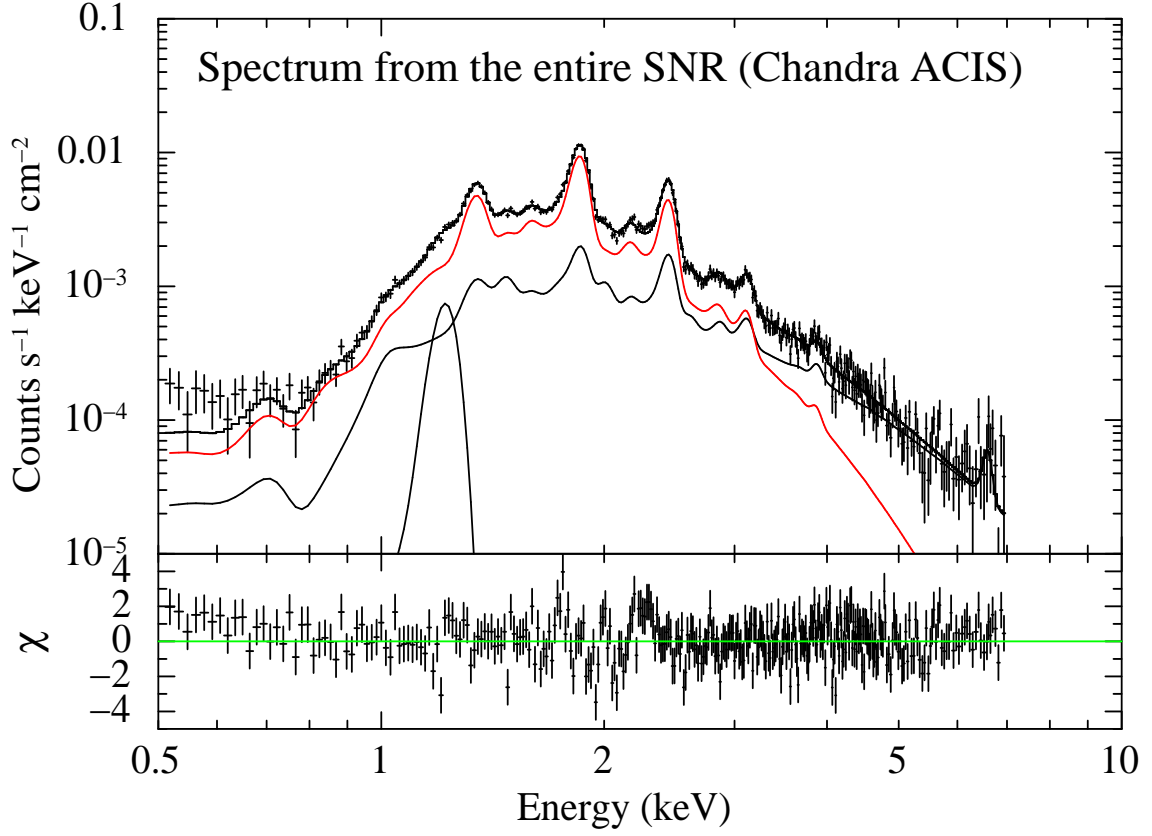}{0.9\textwidth}{}
          }
\caption{Same as Fig.~\ref{fig:kes73_spec}, but for Kes~73. We fit a spectrum from the entire SNR with IME-rich ejecta and CSM components in red and black, respectively. We also added a Gaussian component at $\sim$1.2 keV.
\label{fig:kes73_spec}}
\end{figure}

\begin{figure}[ht!]
\gridline{\fig{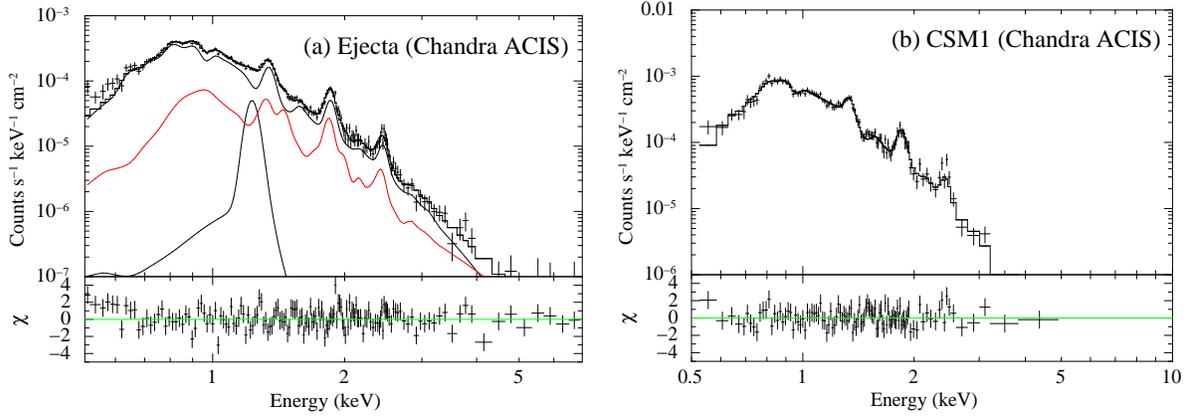}{0.9\textwidth}{}
          }
\caption{Same as Fig.~\ref{fig:casa_spec}, but for RCW~103. 
\label{fig:rcw103_spec}}
\end{figure}

\begin{figure}[ht!]
\gridline{\fig{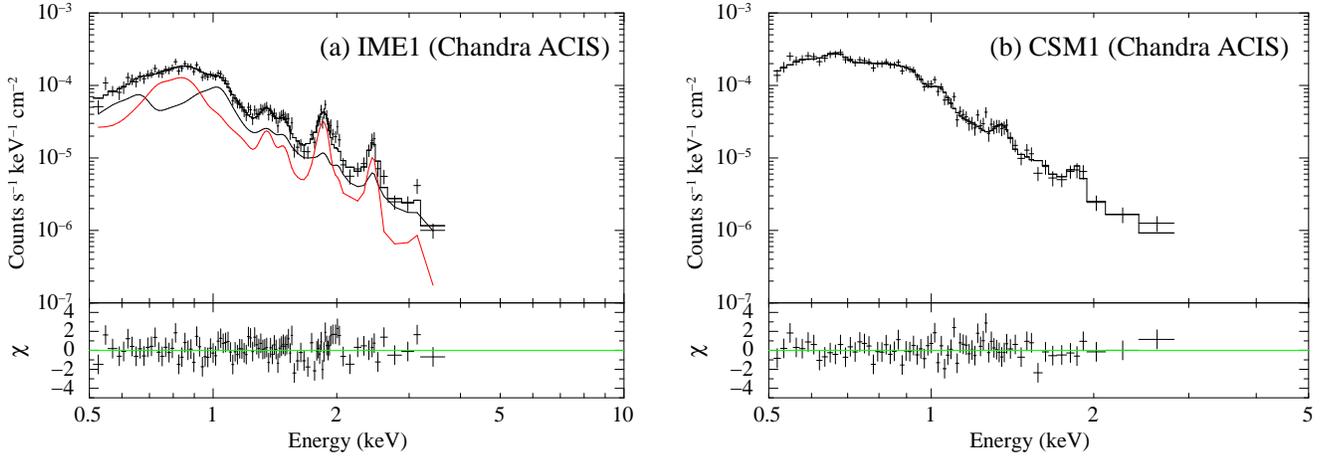}{1.0\textwidth}{}
          }
\caption{Same as Fig.~\ref{fig:casa_spec}, but for N49. (a) Red and black lines represent IME-rich ejecta and CSM components, respectively. (b) We applied a single, CSM component model.
\label{fig:n49_spec}}
\end{figure}

To generate the model spectra for Cas~A, G292.0+1.8, Puppis~A, RCW~103, and N49, we picked up several small regions exhibiting representative X-ray spectra, whereas we used one spectrum extracted from the whole SNR for Kes~73 since its spectral shape is quite uniform everywhere \citep{2014ApJ...781...41K}.  We show these spectral extraction regions in Figure~\ref{fig:specregs}.  For each component (e.g., IME-rich ejecta or O-rich ejecta), we examined three different spectra extracted from different regions (if available) to evaluate systematic uncertainties on the image decomposition and the CoM (see the next section for more details).  Background is taken from off-source regions in the same observations with an exception of the O-rich ejecta region in Puppis~A for which we subtract local background around the feature.  Figures~\ref{fig:casa_spec}--\ref{fig:n49_spec} show representative spectra together with the best-fit models taken from the small or entire regions in the six SNRs, and Tables~\ref{tab:casa_param}--\ref{tab:n49_param} list the best-fit parameters.  The region names in these figures and tables are identical to those in Fig.~\ref{fig:specregs}.

We determined the coefficients $k_j$ in Eq.(1) for each pixel by the following method, and generate $k_j$ distributions as the intensity map of the $j$-th spectral component.  Since the number of photons $n_i$ obeys the Poisson statistical distribution, the most likely $k_j$ is inferred by the maximum likelihood method, that is maximizing 
\begin{equation}
  L = \prod_{i = 1}^{N} \frac{\exp \left[ - f(E_i; k)\right] f(E_i; k)
^{n_i} }{n_i !},
\end{equation}
subject to $k_j \ge 0$.  Solving this problem using standard tool XSPEC may be possible by invoking the process for all pixels ($\sim10^4-10^5$).  This is however not practical due to a limited calculation speed as well as unstable fitting.

Instead, we solve the following approximated problem, that is minimizing $\chi^2_\gamma$ given by 
\begin{equation}
  \chi^2_\gamma = \sum_{i = 1}^N \frac{[ n_i + \min(n_i, 1) - f(E_i; {\bf k})]^2}{n_i + 1},
\end{equation}
subject to $k_j \ge 0$ \citep{1999ApJ...518..380M}.  By subtracting the constant term independent on ${\bf k}$, the problem becomes to minimize 
\begin{equation}
  \widetilde{\chi^2_\gamma} = \sum_{i = 1}^N \left[ \frac{1}{2}
    \left(\sum_{j = 1}^M k_j M_{ji}^{\prime} \right)^2
    - \sum_{j = 1}^M k_j M_{ji}^{\prime\prime} \right],
\end{equation}
subject to $k_j \ge 0$, where $M_{ji}^\prime = \frac{1}{\sqrt{n_i + 1}} M_j(E_i)$ and $M_{ji}^{\prime\prime} = \frac{n_i + \min(n_i, 1)}{n_i + 1} M_j(E_i)$.
By setting $D = M^\prime (M^\prime)^T$, ${\bf b} = {\bf k}$,
${\bf d} = \sum_{i = 1}^N M_{ji}^{\prime\prime}$,
${\bf b_0} = {\bf 0}$, and $A = I$, the problem minimizing $\widetilde{\chi^2_\gamma}$ can be converted to a quadratic programming formulation:
\begin{equation}
  {\rm Minimize} \,\,\,\, \frac{1}{2} {\bf b}^T D {\bf b} - {\bf d}^T {\bf b},
\end{equation}
with the constraints $A^T {\bf b} \ge {\bf b_0}$.
Then, fast computation is possible by using solvers for this type of the problem
.  We used `quadprog' package in R\footnote{http://CRAN.R-project.org/package=quadprog}.

\begin{figure}[ht!]
\gridline{\fig{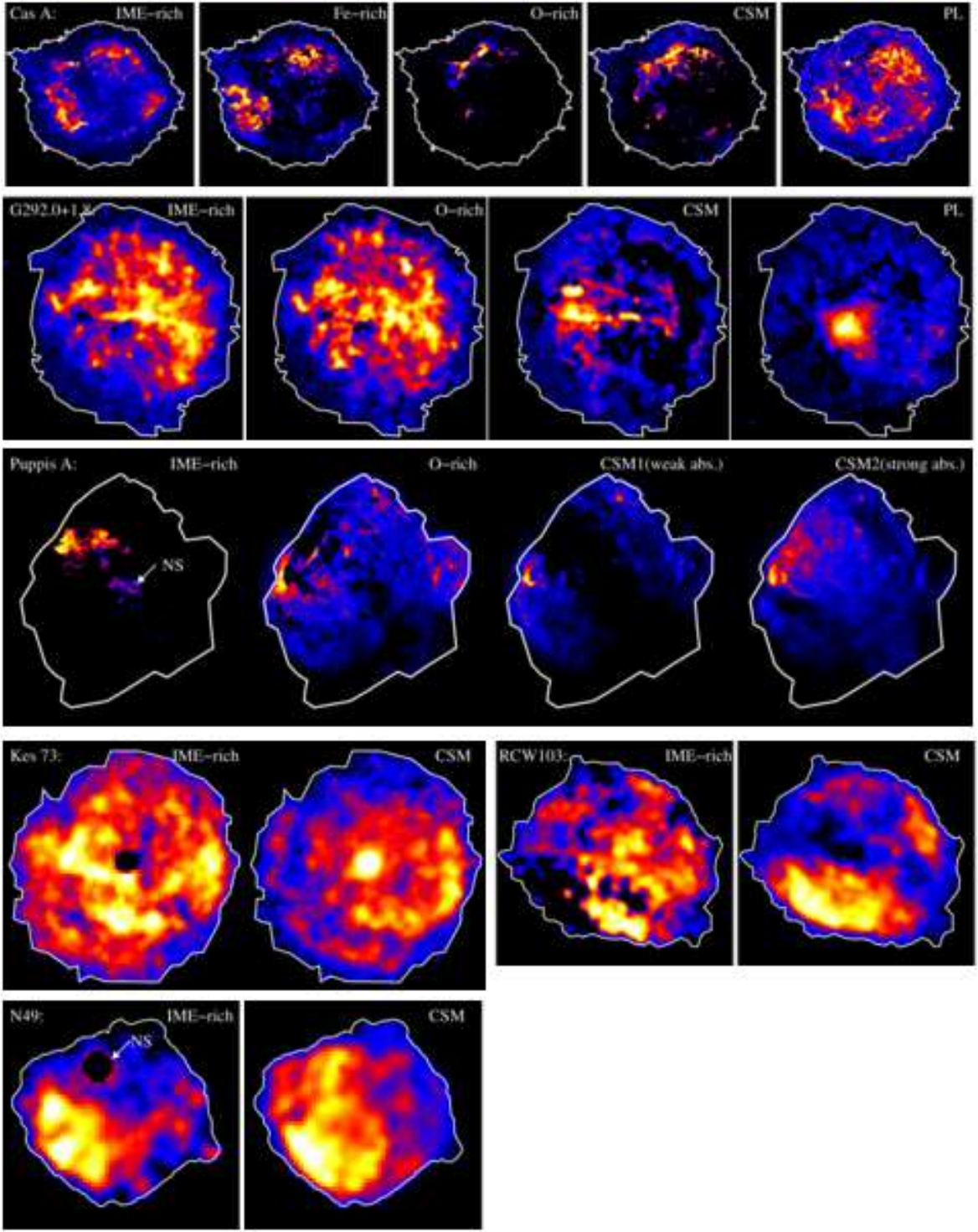}{0.9\textwidth}{}
          }
\caption{Decomposed images for individual spectral model functions for the six SNRs.  The intensity scales are square root. The X-ray boundaries are outlined by white contours. The pixels where we artificially allocate zero values for a robust calculation of the CoMs are indicated by white arrows in the IME-rich ejecta maps in Puppis~A and N49.
\label{fig:decomposed_images}}
\end{figure}

Figure~\ref{fig:decomposed_images} shows the resultant decomposed maps.  The values plotted are products proportional to either $n_{\rm e} n_{\rm H} V$ or the flux at 1\,keV for thermal and power-law components, respectively.  Namely, they are equivalent to normalizations in the models in XSPEC.  We checked the validity of these maps, by comparing them with published results from conventional spatially-resolved spectral mappings that cover the SNRs partially (or fully for Cas~A).  In Cas~A, the IME and Fe maps generally agree with the abundance maps in the literature \citep{2012ApJ...746..130H}.  It should be noted that the well-known IME-rich, jet-like features in the northeast and the southwest are successfully detected in our IME map, although they are not so obvious in our color scale.  In G292.0+1.8, the PL component well traces the pulsar wind nebula \citep{2003ApJ...591L.139H}, and also the O-rich ejecta component's map looks similar to the distribution of optical O-rich fast-moving knots \citep{2009ApJ...692.1489W}.  In Puppis~A, the peak of the IME component coincides with the localized Si and S abundance enhancement in the northeast \citep{2008ApJ...676..378H}, and the O-rich ejecta component's map successfully picks up such ejecta knots/filaments \citep{2008ApJ...678..297K,2010ApJ...714.1725K}.  In Kes~73, the IME component's map generally reflects the X-ray morphology, which is consistent with the fact that X-ray emission is dominated by the ejecta component in this remnant \citep{2014ApJ...781...41K}.  In RCW 103, the IME component's map successfully catches some interior ejecta-dominated features \citep{2015ApJ...810..113F}.  In N49, both the Si-rich shrapnel in the southwest and moderate Si-line enhancements in the southeast \citep{2003ApJ...586..210P} are clearly visible in the IME component's map.   

In addition, we checked goodness of our fittings, by generating maps of residuals between the real images and the sum of individual decomposed images.  We confirmed that the residuals are generally a few 10\% of the data, which would be acceptable given the relatively poor photon statistics in an energy bin of each pixel.  Therefore, we believe that our decomposed maps successfully trace the fundamental properties of the IME-rich ejecta distributions.  

We note in Fig.~\ref{fig:decomposed_images} that NSs are detected in some of the decomposed images, as is evident in the CSM map of Kes~73 in Fig.~\ref{fig:decomposed_images}.  This is caused by the imperfect decomposition at a small fraction of the entire pixels that include emission from NSs.  The same problem can be seen in the IME ejecta maps of Puppis~A and N49, which could affect our measurements of the CoM.  To derive a valid CoM, we artificially set zero values for those ``contaminated" pixels in the IME maps, which are marked by white arrows in Fig.~\ref{fig:decomposed_images}.  

\subsubsection{Computing the CoM}

The values of the IME ejecta maps shown in Fig.~\ref{fig:decomposed_images} are proportional to the emission measure (EM), i.e., $\int n_{\rm e} n_{\rm i} dl$, where $n_{\rm e}$ is the electron density, $n_{\rm i}$ is the ion density, and $dl$ is the plasma depth.  For simplicity, we assume a uniform plasma depth, i.e., 10\% radius, within the entire remnant.  The number of 10\% assumed is never sensitive to the CoM at all, whereas spatial non-uniformity of the plasma depth could affect the estimate of the CoM (cf. the mass is proportional to the square root of the plasma depth).  Then, the square root of the intensity should be proportional to the density and the mass of the X-ray emitting plasma in each pixel.  Therefore, we can estimate a center of mass, by minimizing $\sum_{i = 1}^N m_i L_i$, where $N$ is the number of pixels, $m_i$ is the mass (i.e., the square root of the intensity of the IME component) in the $i$-th pixel, and $L_i$ is the distance between the $i$-th pixel and an arbitrary center of mass.

\begin{figure}[ht!]
\gridline{\fig{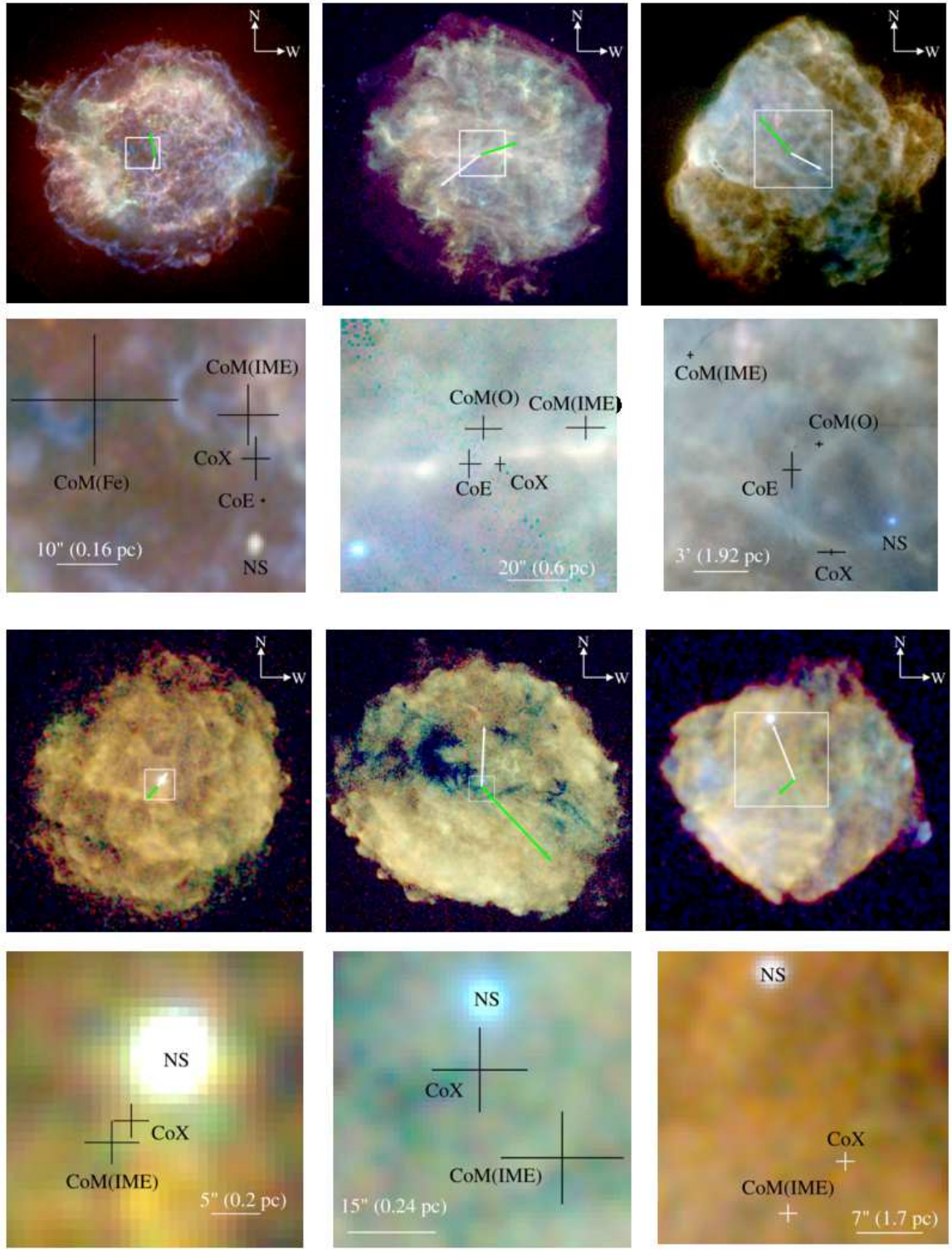}{0.85\textwidth}{}
          }
\caption{Three-color X-ray surface brightness maps in logarithmic scale (upper panels).  Red, green, and blue correspond to energy bands of 0.5--1.5 keV, 1.5--3.0 keV, and 3.0--7.0 keV for Cas~A, 0.5--1.0 keV, 1.0--1.5 keV, and 1.5--3.0 keV for G292.0+1.8, 0.5--0.7 keV, 0.7--1.2 keV, and 1.2--5.0 keV for Puppis~A, 0.5--1.7 keV, 1.7--2.7 keV, and 2.7--7.0 keV for Kes~73, 0.5--1.0 keV, 1.0--1.5 keV, and 1.5--7.0 keV for RCW~103, 0.5--0.8 keV, 0.8--1.7 keV, and 1.7--7.0 keV for N49, respectively.  The white and green arrows point to the direction of NS motion and the center of mass of the IME ejecta, respectively (see also the zoom-up images).  The lengths of the NS vectors are normalized to represent a speed of 1000\,km\,s$^{-1}$ at the best-estimated age and distance (see Table~\ref{tab:targets}).  White boxes around the SNR centers indicate the areas whose close-ups are shown in the lower panels, in which the locations of the center of mass for the IME ejecta (and other elements if available), the CoE if available, and the CoX are overlaid on the X-ray images.  
\label{fig:images}}
\end{figure}

We checked that statistical errors on CoMs are negligibly small (less than the pixel size), by introducing Poisson randomization for the original images and performing the same procedure above.  The uncertainties given in Table~\ref{tab:results} and Fig.~\ref{fig:images} represent systematic errors due to different spectral model functions employed.  We examined three different spectra for each component.  As shown in Fig.~\ref{fig:specregs}, these spectra are extracted from different regions for IME- and Fe-rich components in Cas~A, IME- and O-rich components in G292.0+1.8, and IME-rich component for N49.  For the other components, we artificially generated different spectral models by changing the best-fit parameters of the absorption and the ionization timescale (for the vpshock component) by 10\%, which is comparable with statistical uncertainties on these parameters.   Using different sets of these spectral templates, we calculated dozends / hundreds of CoMs.  Then, we take the average of these CoMs and their standard deviation as our best-estimated CoM and its uncertainty, respectively.  Note that the uncertainties on the asymmetry parameter given in Section~4 are also estimated in the same manner.

\subsection{Estimating the CoX}

\begin{figure}[ht!]
\gridline{\fig{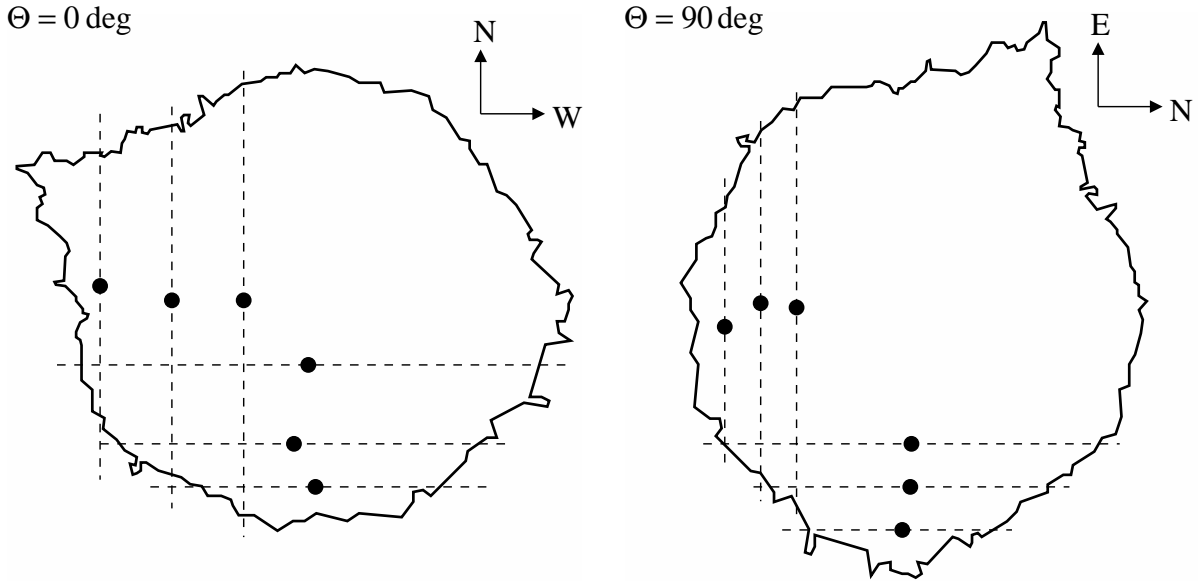}{0.9\textwidth}{}
          }
\caption{X-ray boundary of Cas~A.  To estimate the CoX, we draw horizontal and vertical lines at every pixel (dashed lines for example), and calculate centers of the segments.  Thus-derived centers are averaged, resulting in a single set of x- and y-centers.  By rotating the X-ray boundary (right panel at rotation angle of 90$^\circ$), we repeat the same procedure, obtaining lots of x- and y-centers.  The average and standard deviation of these centers are considered to be the CoX and its uncertainty, respectively.  
\label{fig:casa_edge}}
\end{figure}

To measure the CoX, we first delineated the X-ray boundary of the wide-band (0.5--7\,keV) X-ray image, using contour levels that are sometimes smoothed by eyes if the contour shape is too complicated.  The resultant boundaries are shown as white curves in Fig.~\ref{fig:decomposed_images}.  As shown in Fig.~\ref{fig:casa_edge}, we draw lines on the X-ray boundary along east-west (X) and north-south (Y) directions, and calculated centers of the segments within the SNR boundary in both directions.  We repeat this procedure at every column/row (in units of pixel), resulting in hundreds of centers.  By averaging these centers, we obtained a pair of X and Y centers.  By rotating the X-ray boundary from 0 deg to 180 deg (cf.\ Fig.~\ref{fig:casa_edge} right), we repeat the same procedure, deriving 180 pairs of X and Y centers.  The average and standard deviation of these centers are taken as our best-estimated CoX and its uncertainty, respectively.  The results are given in Table~\ref{tab:results} and Fig.~\ref{fig:images}.  We note that thus-derived CoXs are subject to additional systematic uncertainties in adopting them as the centers of explosion.  This is particularly true if the SNR is interacting with interstellar/molecular clouds.  These include Puppis~A, RCW~103, and N49, as we will describe in the next section.

\section{Results and Discussion}

\begin{figure}[ht!]
\gridline{\fig{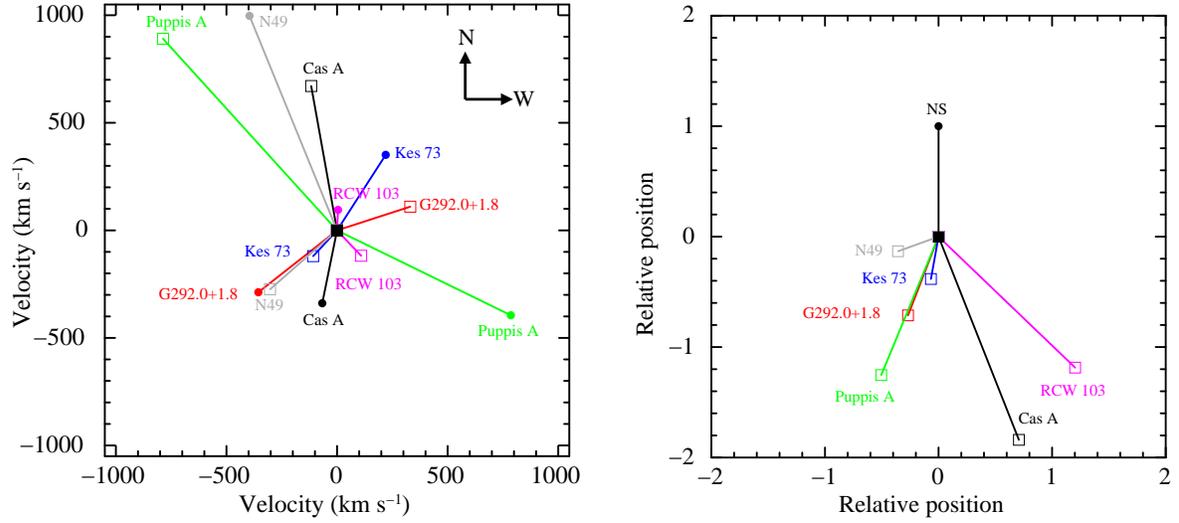}{0.9\textwidth}{}
          }
\caption{Left: NS kick velocities (filled circles) and the CoM velocities (open boxes) with the origin at the CoE or at the CoX for Kes~73, RCW 103, and N49, for which CoEs are not available.  All opening angles between the CoM and the NS are large, which means that CoMs and NSs are located in opposite directions of the explosion points.  The magnetars in Kes~73 and RCW 103 do not possess higher kick velocities than the other NSs.   Right: Same as left but the NS and CoM positions are rotated such that the NS positions are aligned upward, and the velocities are normalized by the NS speeds.
\label{fig:NSrecoil}}
\end{figure}

We found that the NSs and CoMs are located in opposite directions around the CoEs for Cas~A, G292.0+1.8, and Puppis~A, as shown in Table~\ref{tab:results} and Figs.~\ref{fig:images} and \ref{fig:NSrecoil}.  The chance coincidence that we obtain such a good alignment three times is calculated to be only 0.1\%, suggesting that the opposition is not just a coincidence but that there is an underlying physical reason.  Therefore, we argue that the high velocities of young NSs originate from hydrodynamic kicks associated with SN explosion asymmetries.  It is worth noting that the angle of the CoM(IME) around the CoE, i.e., 350$^\circ\pm$20$^\circ$ measured clockwise from celestial north, is consistent with that measured for $^{44}$Ti \citep[340$^\circ\pm$15$^\circ$:][]{2017ApJ...834...19G}.  Given that IMEs in our analysis are dominated by Si and S, we suggest that lighter IMEs are expelled in a similar way to heavier IMEs such as Ti.  The alignment also suggests that IMEs are ejected with a geometry similar to some of the iron peak elements, as Ti is co-produced with Ni in the $\alpha$-rich freezeout.  The match may be imperfect, since multidimensional models for core-collapse SN nucleosynthesis suggest that roughly half of the $^{56}$Ni is produced in the $\alpha$-rich freezeout \citep{2012ApJ...757...69U,2016ApJ...818..123B,2017ApJ...842...13W,2017MNRAS.472..491M}.  Therefore, there is the potential for a second iron peak geometric component from complete Si burning.  However, both production sites ($\alpha$-rich freezeout and explosive Si burning) are so close to each other that they will probably be well mixed (macroscopically) during the later secondary instabilities taking place in SNe, and thus it may be hard to distinguish between the two nucleosynthesis components from observations unless we perform such detailed observations as those for Cas~A.

Additionally, we can see a general opposition between the CoM and the NS pivoted on the CoX for the other three remnants.  Although CoXs may not perfectly point to explosion sites, it should be noted that the CoXs are fairly close ($\sim$2\% of the SNR radius) to the CoEs in Cas~A and G292.0+1.8 (see Table~\ref{tab:results}).  Given that Kes~73 is relatively young and of relatively round shapes like Cas~A and G292.0+1.8, it is reasonable to assume that its CoE is similarly close to the CoXs in these two remnants; 2\% of the SNR radius is comparable with the magnitude of the uncertainty of the CoX.  On the other hand, the displacement between the CoE and the CoX is substantial ($\sim$10\% of the SNR radius) for Puppis~A, which is probably due to the fact that it is a considerably older remnant and is likely to interact with interstellar clouds to the northeast \citep{1995AJ....110..318R}.  N49 is likely to be in the same situation as Puppis~A, since the remnant interacts with an interstellar cloud in the southeast \citep{1997ApJ...480..607B}.  Therefore, the origin of the explosion may be substantially shifted from the CoX to the southeast, which would lead to better alignment for CoM-CoE-NS than that for CoM-CoX-NS, similar to Puppis~A.  RCW~103 is also known to be interacting with a molecular cloud to the southeast \citep{2011AJ....142...47P}.  This implies that the explosion site would be shifted to the south from the CoX.  If we assume the possible displacement to be 10\% of the SNR radius, the explosion site comes close to the east of the CoM(IME), resulting in a worse alignment among CoM, CoE, and NS.  However the relatively round shape of RCW~103 compared with those of Puppis~A and N49 indicates that the displacement between the CoX and the explosion site is not as large as 10\%.  If we assume the displacement to be 2\% of the SNR radius like Cas~A and G292.0+1.8, the CoM-CoE-NS alignment still holds.  

In principle, the ejecta distributions observed in SNRs do not perfectly capture pristine explosion geometries, because SNRs such as our targets are young enough to potentially contain substantial amounts of ``invisible" (namely, cool and thus too dim for us to detect) SN ejecta interior to the reverse shock.  In other words, the distribution of the shocked ejecta is significantly different from the original explosion geometry, if only a tiny portion of the ejecta is heated by the reverse shock and the environmental density structure is highly asymmetric.  However, there are arguments why the masses of such invisible ejecta are likely to be much smaller than those of the shock-heated ejecta in all of the six SNRs of our interest.  Extensive studies of Cas~A showed that there is only little mass of unshocked ejecta remaining in the central volume; the total amount of the unshocked ejecta was estimated to be only $\sim$0.3 M$_\odot$, corresponding to $\sim$10\% of the shock-heated ejecta \citep{2012ApJ...746..130H,2014ApJ...785....7D}.  The unshocked ejecta are considered to be dominated by O and Si, based on the analysis of infrared lines of [Si II], [O IV], [S III], and [S IV] \citep{2014ApJ...785....7D}.  As in a previous study \citep{2014ApJ...785....7D}, we rely here on X-ray derived abundances [O : Ne : Mg : Si : S : Ar : Fe = 1 : 0.015 : 0.004 : 0.021 : 0.011 : 0.0056 : 0.054 in mass ratios measured by \citet{2012ApJ...746..130H}]. We find that the fractional IME mass in the unshocked ejecta (of $\sim$0.3\,M$_\odot$) is only $\sim$4\% or $\sim$0.01\,M$_\odot$. This is an order of magnitude smaller than the mass of the shock-heated IMEs ($\sim$0.06--0.14\,M$_\odot$, depending on the plasma depth and the metallicity).  Cas~A is likely to have the most unshocked ejecta given its youth.  In fact, the evolutionary stages for the other five remnants are greater than or similar to that of Cas~A (see Table~\ref{tab:targets}).  It is thus reasonable to assume that the unshocked ejecta masses in all the other remnants are smaller than or at least in an order-of-magnitude agreement with that in Cas~A.  Therefore the masses of unshocked IMEs can be expected to be significantly smaller than those of the shock-heated IMEs in Table~\ref{tab:results}. For these reasons, we believe that the IME distributions shown in Fig.~\ref{fig:decomposed_images} are good representations of the total IME masses.

If NSs are kicked by the hydrodynamical mechanism, then the NS kick velocities should correlate with the momentum-asymmetry parameter, $\alpha_{\rm ej}$, defined as $\int dV \rho v / \int dV \rho |v|$ with $V$ being the volume, $\rho$ the density, and $v$ the velocity of the relevant ejecta to be integrated, as well as the explosion energy and the NS mass \citep[see Eq.(11) in][]{2017ApJ...837...84J}.  One expects a correlation with a considerable scatter between the NS velocity and the $\alpha_{\rm ej}$ parameter, because NS masses can vary by several 10\% and SN energies are typically constrained within a range of some 10$^{50}$ erg and about 2$\times$10$^{51}$ erg.  

\begin{figure}[ht!]
\gridline{\fig{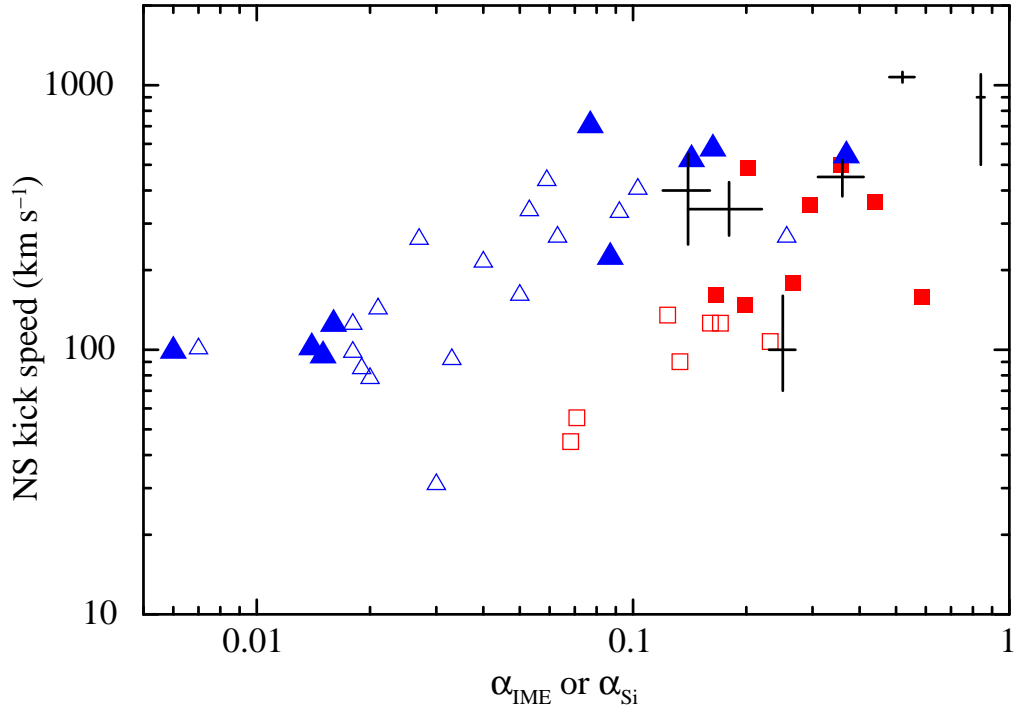}{0.8\textwidth}{}
          }
\caption{NS kick velocities versus asymmetry parameters for the IMEs.  Observational values are shown as crosses.  Results from numerical 2D and 3D simulations \citep{2016MNRAS.461.3296N,2013A&A...552A.126W} are shown as squares and triangles, respectively.  They are calculated for Si interior to the shock radius at a time of about 1 second (open symbols) and about 3 seconds (filled symbols) after core bounce.
\label{fig:kick_vs_alpha}}
\end{figure}

We thus compared $\alpha_{\rm ej}$ measured in SNRs with those based on recent state-of-the-art 2D and 3D hydrodynamic simulations of neutrino-driven explosions \citep{2006A&A...457..963S,2013A&A...552A.126W,2017ApJ...837...84J}.  In order to estimate the value of $\alpha_{\rm IME}$ for the distribution of IME ejecta in SNRs, we integrated the product of the square root of the intensity of the IME component (which is a proxy of the density and mass) and the distance from the CoE/CoX (which is a proxy for the velocity) over all pixels.  As for the simulated values, the asymmetry parameters listed in the literature for SN models \citep{2006A&A...457..963S,2013A&A...552A.126W} are measured for the total expelled mass behind the shock at about the time when the NS kick is determined (because this mass is dynamically relevant for the NS acceleration). In contrast, in order to facilitate a better comparison with our observations, which focus on the IMEs, we computed the simulated asymmetry parameters for Si in the present work.  The resultant values of $\alpha_{\rm IME}$ (observations) and $\alpha_{\rm Si}$ (simulations) are shown in Fig.~\ref{fig:kick_vs_alpha}.  Indeed, we do find a positive correlation between observed kick velocities and $\alpha_{\rm IME}$, in qualitative agreement with the numerical SN simulations (Fig.~\ref{fig:kick_vs_alpha}), despite the fact that the considered evolution stages differ in time by some 10 orders of magnitude ($\sim$300--5000 yrs compared to $\sim$1 second).

The observational values of $\alpha_{\rm IME}$, however, exhibit a tendency to be larger (0.16--0.84) than $\alpha_{\rm Si}$ from the simulations ($\lesssim$0.5). An overestimation of the observationally determined values of $\alpha_{\rm IME}$ caused by an imperfect decomposition of SN ejecta and CSM, and by only incomplete heating of the IME ejecta by the reverse shock, is unlikely to explain this difference (see above). Instead, there are several reasons connected to the theoretical data. First, $\alpha_{\rm Si}$ is on the low side as a proxy of $\alpha_{\rm IME}$, because elements such as S and Ar (which are not available for some of the models) show larger asymmetries than Si.  

Second, a significant part of the explosion simulations was evolved only for slightly more than one second, whereas both the NS kick velocity and αSi can increase over several seconds. The cause for the long-time growth of αSi is two-fold. On one hand, IME (Si) nucleosynthesis can continue for more than one second, in particular in the hemisphere of the stronger explosion. On the other hand, the internal thermal energy deposited by the explosion mechanism (neutrino heating in the computed models) gets converted to kinetic energy only over time scales much longer than a few seconds. Therefore the nucleosynthetic products will experience an ongoing acceleration, which is also larger in the hemisphere of the stronger explosion, enhancing the long-time growth of $\alpha_{\rm Si}$.  In fact, this effect can be seen as a systematic increase of αSi with time for models whose values are displayed both at an early ($\sim$1 second) and at a later ($\sim$3 seconds) evolution stage in Fig.~\ref{fig:kick_vs_alpha}.  

Third, accounting for projection effects instead of our 3D evaluation of the numerical models \citep{2006A&A...457..963S,2013A&A...552A.126W} may also lead to smaller tangential NS velocities combined with larger values of $\alpha_{\rm Si}$, depending on the viewing angle and the explosion anisotropy.  This tends to move the theoretical points toward the lower-right corner of Fig.~\ref{fig:kick_vs_alpha}, thus improving the agreement with the observations.  Fourth and final, the available sets of 2D and 3D explosion models are based only on a limited range of progenitor stars and explosion conditions, which are statistically not representative of all possible variations over the full mass spectrum of SN progenitors. In particular, cases that produce very high NS kicks ($>$1000 km\,s$^{-1}$) are absent in our current sample of SN models.  

The opposite directions of dominant IME ejection and NS motion, together with the discovered correlation between NS kick velocities and $\alpha_{\rm IME}$, strengthen the case for the hydrodynamic NS kick scenario developed by the recent core-collapse SN simulations.  This also supports global explosion asphericities being mainly caused by various kinds of hydrodynamic instabilities such as neutrino-driven convection \citep{1994ApJ...435..339H,1995ApJ...450..830B,1996A&A...306..167J} and the standing-accretion shock instability \citep{2003ApJ...584..971B}.

A large number of studies have investigated the question whether or not NSs with greater velocities possess stronger surface dipole-magnetic fields \citep[e.g.,][]{1995MNRAS.275L..16L}.  In our sample, the two magnetars in Kes~73 and RCW~103 have relatively low kick velocities compared to three NSs that are called ``central compact objects" and whose magnetic fields are thought to be low.  Also the measured velocities of other young magnetars [$\sim$210 km\,s$^{-1}$ for XTE J1810-197 \citep{2007ApJ...662.1198H}; $\sim$280 km\,s$^{-1}$ for PSR J1550-5418 \citep{2012ApJ...748L...1D}; $\sim$350 km\,s$^{-1}$ for SGR 1806-20 \citep{2012ApJ...761...76T}; $\sim$130 km\,s$^{-1}$ for SGR 1900+14 \citep{2012ApJ...761...76T}] do not yield evidence that magnetars have higher kicks than normal NSs.  This fact argues against neutrino-induced kick scenarios in which the NS velocities are expected to correlate positively with the magnetic-field strength \citep{1993AAT....3..287B,2005ApJ...632..531S,2003PhRvD..68j3002F,2008A&A...489..281S,2012PhRvD..86l3003M,2006ApJS..163..335F}.  The possible alignment between spin and velocity orientations for many NSs, which has been inferred from X-ray imaging of pulsar wind nebulae \citep{2007ApJ...660.1357N} and polarimetric observations of radio pulsars \citep[e.g.,][for a counter argument]{2013MNRAS.430.2281N,2007MNRAS.381.1625J}, does not provide a strong support for neutrino-induced NS kicks directed by strong magnetic fields either, because spin-kick alignment might also be a natural consequence of the hydrodynamic kick scenario \citep{2017ApJ...837...84J,2017MNRAS.472..491M}.  Moreover, there are several pulsar wind nebulae showing clear evidence against spin-kick alignment: G292.0+1.8 \citep{2007ApJ...670L.121P}; Geminga \citep{2017ApJ...835...66P}; IGR~J11014-6103 \citep{2014A&A...562A.122P}; Guitar \citep{2012ApJ...747...74H}; 3C58 \citep{2013MNRAS.431.2590B}.  We also note that a recent study of SNRs ruled out jet-kick alignment \citep{2017arXiv171000819B}, which is in tension with the spin-kick alignment.  More work is needed on the question for spin-kick alignment both on the observational and theoretical sides.

\section{Conclusions} \label{sec:conclusions}

We revealed X-ray emitting IME-rich ejecta distributions for six young SNRs, based on an image decomposition technique that we developed for this work.  We found that the centers of IME-ejecta masses are shifted from the explosion sites in the direction opposite to the NS's apparent motion for all the six SNRs.  We also found that the NS kick speeds correlate with the degree of asymmetries of the IME-rich ejecta.  Additionally, there is no correlation between the kick velocities and magnetic field strengths of these neutron stars, as has been previously reported.  These results are fully consistent with the hydrodynamic-kick scenario rather than the neutrino-induced kick scenario, and supports that global explosion asphericities are mainly caused by various kinds of hydrodynamic instabilities such as neutrino-driven convection.  Our work established a long-suspected link between the SN asymmetries and NS kicks.

Our result is generally consistent with the recent work by \citet{2017ApJ...844...84H}.  Four objects are shared between the two analyses: Cas~A, G292.0+1.8, Puppis~A, and RCW~103.  Of these, Cas~A, G292.0+1.8, and Puppis~A show the same results.  However, we found an inconsistent result for RCW~103.  Another difference is that \citet{2017ApJ...844...84H} found no correlation between the degree of asymmetries (dipole, quadrupole, or octupole) and the NS kick velocities, whereas we obtained a positive correlation.  Given that our work has an advantage that separates the CSM and ejecta components, we believe that our results are more reliable than the previous ones, and finally provides us with conclusive evidence for the hydrodynamic NS kick mechanism.

\acknowledgments

This report is based on archival data acquired with the X-ray observatories {\it Chandra} and {\it XMM-Newton}.  Numerical computations were carried out on the Cray XC30 at the Center for Computational Astrophysics, National Astronomical Observatory of Japan.  We are grateful to S.\ Ikeda for advisements on the statistical treatment of the data analysis, and P.F.\ Winkler for providing us with proper-motion data of O-rich knots in Puppis~A and G292.0+1.8.  This work was supported by the Japan Society for the Promotion of Science KAKENHI grant numbers 16K17673, 17H02864 (SK), 17K05395 (MM), JP15KK0173, JP17H06364, JP17H01130 (KK), 16H03983 (KM), 15H02075 (MT), and in Garching by the Deutsche Forschungsgemeinschaft through the Excellence Cluster ``Universe" EXC 153 and by the European Research Council through grant ERC-AdG No. 341157-COCO2CASA.  This work was also supported by JICFuS as a priority issue to be tackled by using the Post `K' Computer, and by CREST, Japan Science and Technology Agency (JST).  This work was partly supported by Leading Initiative for Excellent Young Researchers, MEXT, Japan.  We also thank anonymous referees for their constructive comments that improved the quality of the paper.


\begin{thebibliography}{}
\expandafter\ifx\csname natexlab\endcsname\relax\def\natexlab#1{#1}\fi

\bibitem[{{Arnaud}(1996)}]{1996ASPC..101...17A}
{Arnaud}, K.~A. 1996, in Astronomical Society of the Pacific Conference Series,
  Vol. 101, Astronomical Data Analysis Software and Systems V, ed. G.~H.
  {Jacoby} \& J.~{Barnes}, 17

\bibitem[{{Banas} {et~al.}(1997){Banas}, {Hughes}, {Bronfman}, \&
  {Nyman}}]{1997ApJ...480..607B}
{Banas}, K.~R., {Hughes}, J.~P., {Bronfman}, L., \& {Nyman}, L.-{\AA}. 1997,
  \apj, 480, 607

\bibitem[{{Bear} \& {Soker}(2017)}]{2017arXiv171000819B}
{Bear}, E., \& {Soker}, N. 2017, ArXiv e-prints, arXiv:1710.00819

\bibitem[{{Becker} {et~al.}(2012){Becker}, {Prinz}, {Winkler}, \&
  {Petre}}]{2012ApJ...755..141B}
{Becker}, W., {Prinz}, T., {Winkler}, P.~F., \& {Petre}, R. 2012, \apj, 755,
  141

\bibitem[{{Bhalerao} {et~al.}(2015){Bhalerao}, {Park}, {Dewey}, {Hughes},
  {Mori}, \& {Lee}}]{2015ApJ...800...65B}
{Bhalerao}, J., {Park}, S., {Dewey}, D., {et~al.} 2015, \apj, 800, 65

\bibitem[{{Bietenholz} {et~al.}(2013){Bietenholz}, {Kondratiev}, {Ransom},
  {Slane}, {Bartel}, \& {Buchner}}]{2013MNRAS.431.2590B}
{Bietenholz}, M.~F., {Kondratiev}, V., {Ransom}, S., {et~al.} 2013, \mnras,
  431, 2590

\bibitem[{{Bisnovatyi-Kogan}(1993)}]{1993AAT....3..287B}
{Bisnovatyi-Kogan}, G.~S. 1993, Astronomical and Astrophysical Transactions, 3,
  287

\bibitem[{{Blaauw}(1961)}]{1961BAN....15..265B}
{Blaauw}, A. 1961, \bain, 15, 265

\bibitem[{{Blondin} {et~al.}(2003){Blondin}, {Mezzacappa}, \&
  {DeMarino}}]{2003ApJ...584..971B}
{Blondin}, J.~M., {Mezzacappa}, A., \& {DeMarino}, C. 2003, \apj, 584, 971

\bibitem[{{Borkowski} {et~al.}(2001){Borkowski}, {Lyerly}, \&
  {Reynolds}}]{2001ApJ...548..820B}
{Borkowski}, K.~J., {Lyerly}, W.~J., \& {Reynolds}, S.~P. 2001, \apj, 548, 820

\bibitem[{{Bruenn} {et~al.}(2016){Bruenn}, {Lentz}, {Hix}, {Mezzacappa},
  {Harris}, {Messer}, {Endeve}, {Blondin}, {Chertkow}, {Lingerfelt},
  {Marronetti}, \& {Yakunin}}]{2016ApJ...818..123B}
{Bruenn}, S.~W., {Lentz}, E.~J., {Hix}, W.~R., {et~al.} 2016, \apj, 818, 123

\bibitem[{{Burrows} \& {Hayes}(1996)}]{1996PhRvL..76..352B}
{Burrows}, A., \& {Hayes}, J. 1996, Physical Review Letters, 76, 352

\bibitem[{{Burrows} {et~al.}(1995){Burrows}, {Hayes}, \&
  {Fryxell}}]{1995ApJ...450..830B}
{Burrows}, A., {Hayes}, J., \& {Fryxell}, B.~A. 1995, \apj, 450, 830

\bibitem[{{Carter} {et~al.}(1997){Carter}, {Dickel}, \&
  {Bomans}}]{1997PASP..109..990C}
{Carter}, L.~M., {Dickel}, J.~R., \& {Bomans}, D.~J. 1997, \pasp, 109, 990

\bibitem[{{Chevalier} \& {Oishi}(2003)}]{2003ApJ...593L..23C}
{Chevalier}, R.~A., \& {Oishi}, J. 2003, \apjl, 593, L23

\bibitem[{{De Luca} {et~al.}(2006){De Luca}, {Caraveo}, {Mereghetti}, {Tiengo},
  \& {Bignami}}]{2006Sci...313..814D}
{De Luca}, A., {Caraveo}, P.~A., {Mereghetti}, S., {Tiengo}, A., \& {Bignami},
  G.~F. 2006, Science, 313, 814

\bibitem[{{DeLaney} {et~al.}(2014){DeLaney}, {Kassim}, {Rudnick}, \&
  {Perley}}]{2014ApJ...785....7D}
{DeLaney}, T., {Kassim}, N.~E., {Rudnick}, L., \& {Perley}, R.~A. 2014, \apj,
  785, 7

\bibitem[{{Deller} {et~al.}(2012){Deller}, {Camilo}, {Reynolds}, \&
  {Halpern}}]{2012ApJ...748L...1D}
{Deller}, A.~T., {Camilo}, F., {Reynolds}, J.~E., \& {Halpern}, J.~P. 2012,
  \apjl, 748, L1

\bibitem[{{Esposito} {et~al.}(2011){Esposito}, {Turolla}, {de Luca}, {Israel},
  {Possenti}, \& {Burrows}}]{2011MNRAS.418..170E}
{Esposito}, P., {Turolla}, R., {de Luca}, A., {et~al.} 2011, \mnras, 418, 170

\bibitem[{{Feast}(1999)}]{1999PASP..111..775F}
{Feast}, M. 1999, \pasp, 111, 775

\bibitem[{{Fesen} {et~al.}(2006){Fesen}, {Hammell}, {Morse}, {Chevalier},
  {Borkowski}, {Dopita}, {Gerardy}, {Lawrence}, {Raymond}, \& {van den
  Bergh}}]{2006ApJ...645..283F}
{Fesen}, R.~A., {Hammell}, M.~C., {Morse}, J., {et~al.} 2006, \apj, 645, 283

\bibitem[{{Frank} {et~al.}(2015){Frank}, {Burrows}, \&
  {Park}}]{2015ApJ...810..113F}
{Frank}, K.~A., {Burrows}, D.~N., \& {Park}, S. 2015, \apj, 810, 113

\bibitem[{{Fryer} \& {Kusenko}(2006)}]{2006ApJS..163..335F}
{Fryer}, C.~L., \& {Kusenko}, A. 2006, \apjs, 163, 335

\bibitem[{{Fuller} {et~al.}(2003){Fuller}, {Kusenko}, {Mocioiu}, \&
  {Pascoli}}]{2003PhRvD..68j3002F}
{Fuller}, G.~M., {Kusenko}, A., {Mocioiu}, I., \& {Pascoli}, S. 2003, \prd, 68,
  103002

\bibitem[{{Gaensler} \& {Wallace}(2003)}]{2003ApJ...594..326G}
{Gaensler}, B.~M., \& {Wallace}, B.~J. 2003, \apj, 594, 326

\bibitem[{{Garmire} {et~al.}(2000){Garmire}, {Pavlov}, {Garmire}, \&
  {Zavlin}}]{2000IAUC.7350....2G}
{Garmire}, G.~P., {Pavlov}, G.~G., {Garmire}, A.~B., \& {Zavlin}, V.~E. 2000,
  \iaucirc, 7350

\bibitem[{{Gotthelf} \& {Halpern}(2009)}]{2009ApJ...695L..35G}
{Gotthelf}, E.~V., \& {Halpern}, J.~P. 2009, \apjl, 695, L35

\bibitem[{{Gotthelf} {et~al.}(2013){Gotthelf}, {Halpern}, \&
  {Alford}}]{2013ApJ...765...58G}
{Gotthelf}, E.~V., {Halpern}, J.~P., \& {Alford}, J. 2013, \apj, 765, 58

\bibitem[{{Gotthelf} {et~al.}(2001){Gotthelf}, {Koralesky}, {Rudnick}, {Jones},
  {Hwang}, \& {Petre}}]{2001ApJ...552L..39G}
{Gotthelf}, E.~V., {Koralesky}, B., {Rudnick}, L., {et~al.} 2001, \apjl, 552,
  L39

\bibitem[{{Grefenstette} {et~al.}(2014){Grefenstette}, {Harrison}, {Boggs},
  {Reynolds}, {Fryer}, {Madsen}, {Wik}, {Zoglauer}, {Ellinger}, {Alexander},
  {An}, {Barret}, {Christensen}, {Craig}, {Forster}, {Giommi}, {Hailey},
  {Hornstrup}, {Kaspi}, {Kitaguchi}, {Koglin}, {Mao}, {Miyasaka}, {Mori},
  {Perri}, {Pivovaroff}, {Puccetti}, {Rana}, {Stern}, {Westergaard}, \&
  {Zhang}}]{2014Natur.506..339G}
{Grefenstette}, B.~W., {Harrison}, F.~A., {Boggs}, S.~E., {et~al.} 2014, \nat,
  506, 339

\bibitem[{{Grefenstette} {et~al.}(2017){Grefenstette}, {Fryer}, {Harrison},
  {Boggs}, {DeLaney}, {Laming}, {Reynolds}, {Alexander}, {Barret},
  {Christensen}, {Craig}, {Forster}, {Giommi}, {Hailey}, {Hornstrup},
  {Kitaguchi}, {Koglin}, {Lopez}, {Mao}, {Madsen}, {Miyasaka}, {Mori}, {Perri},
  {Pivovaroff}, {Puccetti}, {Rana}, {Stern}, {Westergaard}, {Wik}, {Zhang}, \&
  {Zoglauer}}]{2017ApJ...834...19G}
{Grefenstette}, B.~W., {Fryer}, C.~L., {Harrison}, F.~A., {et~al.} 2017, \apj,
  834, 19

\bibitem[{{Helfand} {et~al.}(2007){Helfand}, {Chatterjee}, {Brisken}, {Camilo},
  {Reynolds}, {van Kerkwijk}, {Halpern}, \& {Ransom}}]{2007ApJ...662.1198H}
{Helfand}, D.~J., {Chatterjee}, S., {Brisken}, W.~F., {et~al.} 2007, \apj, 662,
  1198

\bibitem[{{Herant} {et~al.}(1994){Herant}, {Benz}, {Hix}, {Fryer}, \&
  {Colgate}}]{1994ApJ...435..339H}
{Herant}, M., {Benz}, W., {Hix}, W.~R., {Fryer}, C.~L., \& {Colgate}, S.~A.
  1994, \apj, 435, 339

\bibitem[{{Hobbs} {et~al.}(2005){Hobbs}, {Lorimer}, {Lyne}, \&
  {Kramer}}]{2005MNRAS.360..974H}
{Hobbs}, G., {Lorimer}, D.~R., {Lyne}, A.~G., \& {Kramer}, M. 2005, \mnras,
  360, 974

\bibitem[{{Holland-Ashford} {et~al.}(2017){Holland-Ashford}, {Lopez},
  {Auchettl}, {Temim}, \& {Ramirez-Ruiz}}]{2017ApJ...844...84H}
{Holland-Ashford}, T., {Lopez}, L.~A., {Auchettl}, K., {Temim}, T., \&
  {Ramirez-Ruiz}, E. 2017, \apj, 844, 84

\bibitem[{{Hughes} {et~al.}(1998){Hughes}, {Hayashi}, \&
  {Koyama}}]{1998ApJ...505..732H}
{Hughes}, J.~P., {Hayashi}, I., \& {Koyama}, K. 1998, \apj, 505, 732

\bibitem[{{Hughes} {et~al.}(2003){Hughes}, {Slane}, {Park}, {Roming}, \&
  {Burrows}}]{2003ApJ...591L.139H}
{Hughes}, J.~P., {Slane}, P.~O., {Park}, S., {Roming}, P.~W.~A., \& {Burrows},
  D.~N. 2003, \apjl, 591, L139

\bibitem[{{Hui} {et~al.}(2012){Hui}, {Huang}, {Trepl}, {Tetzlaff}, {Takata},
  {Wu}, \& {Cheng}}]{2012ApJ...747...74H}
{Hui}, C.~Y., {Huang}, R.~H.~H., {Trepl}, L., {et~al.} 2012, \apj, 747, 74

\bibitem[{{Hwang} \& {Laming}(2012)}]{2012ApJ...746..130H}
{Hwang}, U., \& {Laming}, J.~M. 2012, \apj, 746, 130

\bibitem[{{Hwang} {et~al.}(2008){Hwang}, {Petre}, \&
  {Flanagan}}]{2008ApJ...676..378H}
{Hwang}, U., {Petre}, R., \& {Flanagan}, K.~A. 2008, \apj, 676, 378

\bibitem[{{Janka}(2017)}]{2017ApJ...837...84J}
{Janka}, H.-T. 2017, \apj, 837, 84

\bibitem[{{Janka} \& {Mueller}(1994)}]{1994A&A...290..496J}
{Janka}, H.-T., \& {Mueller}, E. 1994, \aap, 290, 496

\bibitem[{{Janka} \& {Mueller}(1996)}]{1996A&A...306..167J}
---. 1996, \aap, 306, 167

\bibitem[{{Johnston} {et~al.}(2007){Johnston}, {Kramer}, {Karastergiou},
  {Hobbs}, {Ord}, \& {Wallman}}]{2007MNRAS.381.1625J}
{Johnston}, S., {Kramer}, M., {Karastergiou}, A., {et~al.} 2007, \mnras, 381,
  1625

\bibitem[{{Katsuda} {et~al.}(2010{\natexlab{a}}){Katsuda}, {Hwang}, {Petre},
  {Park}, {Mori}, \& {Tsunemi}}]{2010ApJ...714.1725K}
{Katsuda}, S., {Hwang}, U., {Petre}, R., {et~al.} 2010{\natexlab{a}}, \apj,
  714, 1725

\bibitem[{{Katsuda} {et~al.}(2008){Katsuda}, {Mori}, {Tsunemi}, {Park},
  {Hwang}, {Burrows}, {Hughes}, \& {Slane}}]{2008ApJ...678..297K}
{Katsuda}, S., {Mori}, K., {Tsunemi}, H., {et~al.} 2008, \apj, 678, 297

\bibitem[{{Katsuda} {et~al.}(2013){Katsuda}, {Ohira}, {Mori}, {Tsunemi},
  {Uchida}, {Koyama}, \& {Tamagawa}}]{2013ApJ...768..182K}
{Katsuda}, S., {Ohira}, Y., {Mori}, K., {et~al.} 2013, \apj, 768, 182

\bibitem[{{Katsuda} {et~al.}(2010{\natexlab{b}}){Katsuda}, {Petre}, {Hughes},
  {Hwang}, {Yamaguchi}, {Hayato}, {Mori}, \& {Tsunemi}}]{2010ApJ...709.1387K}
{Katsuda}, S., {Petre}, R., {Hughes}, J.~P., {et~al.} 2010{\natexlab{b}}, \apj,
  709, 1387

\bibitem[{{Kulkarni} {et~al.}(2003){Kulkarni}, {Kaplan}, {Marshall}, {Frail},
  {Murakami}, \& {Yonetoku}}]{2003ApJ...585..948K}
{Kulkarni}, S.~R., {Kaplan}, D.~L., {Marshall}, H.~L., {et~al.} 2003, \apj,
  585, 948

\bibitem[{{Kumar} {et~al.}(2014){Kumar}, {Safi-Harb}, {Slane}, \&
  {Gotthelf}}]{2014ApJ...781...41K}
{Kumar}, H.~S., {Safi-Harb}, S., {Slane}, P.~O., \& {Gotthelf}, E.~V. 2014,
  \apj, 781, 41

\bibitem[{{Lai} {et~al.}(2001){Lai}, {Chernoff}, \&
  {Cordes}}]{2001ApJ...549.1111L}
{Lai}, D., {Chernoff}, D.~F., \& {Cordes}, J.~M. 2001, \apj, 549, 1111

\bibitem[{{Leonard} {et~al.}(2006){Leonard}, {Filippenko}, {Ganeshalingam},
  {Serduke}, {Li}, {Swift}, {Gal-Yam}, {Foley}, {Fox}, {Park}, {Hoffman}, \&
  {Wong}}]{2006Natur.440..505L}
{Leonard}, D.~C., {Filippenko}, A.~V., {Ganeshalingam}, M., {et~al.} 2006,
  \nat, 440, 505

\bibitem[{{Lorimer} {et~al.}(1995){Lorimer}, {Lyne}, \&
  {Anderson}}]{1995MNRAS.275L..16L}
{Lorimer}, D.~R., {Lyne}, A.~G., \& {Anderson}, B. 1995, \mnras, 275, L16

\bibitem[{{Maeda} {et~al.}(2008){Maeda}, {Kawabata}, {Mazzali}, {Tanaka},
  {Valenti}, {Nomoto}, {Hattori}, {Deng}, {Pian}, {Taubenberger}, {Iye},
  {Matheson}, {Filippenko}, {Aoki}, {Kosugi}, {Ohyama}, {Sasaki}, \&
  {Takata}}]{2008Sci...319.1220M}
{Maeda}, K., {Kawabata}, K., {Mazzali}, P.~A., {et~al.} 2008, Science, 319,
  1220

\bibitem[{{Maruyama} {et~al.}(2012){Maruyama}, {Yasutake}, {Cheoun}, {Hidaka},
  {Kajino}, {Mathews}, \& {Ryu}}]{2012PhRvD..86l3003M}
{Maruyama}, T., {Yasutake}, N., {Cheoun}, M.-K., {et~al.} 2012, \prd, 86,
  123003

\bibitem[{{Mighell}(1999)}]{1999ApJ...518..380M}
{Mighell}, K.~J. 1999, \apj, 518, 380

\bibitem[{{M{\"u}ller} {et~al.}(2017){M{\"u}ller}, {Melson}, {Heger}, \&
  {Janka}}]{2017MNRAS.472..491M}
{M{\"u}ller}, B., {Melson}, T., {Heger}, A., \& {Janka}, H.-T. 2017, \mnras,
  472, 491

\bibitem[{{Nagataki}(2000)}]{2000ApJS..127..141N}
{Nagataki}, S. 2000, \apjs, 127, 141

\bibitem[{{Nakamura} {et~al.}(2016){Nakamura}, {Horiuchi}, {Tanaka}, {Hayama},
  {Takiwaki}, \& {Kotake}}]{2016MNRAS.461.3296N}
{Nakamura}, K., {Horiuchi}, S., {Tanaka}, M., {et~al.} 2016, \mnras, 461, 3296

\bibitem[{{Ng} \& {Romani}(2007)}]{2007ApJ...660.1357N}
{Ng}, C.-Y., \& {Romani}, R.~W. 2007, \apj, 660, 1357

\bibitem[{{Nordhaus} {et~al.}(2012){Nordhaus}, {Brandt}, {Burrows}, \&
  {Almgren}}]{2012MNRAS.423.1805N}
{Nordhaus}, J., {Brandt}, T.~D., {Burrows}, A., \& {Almgren}, A. 2012, \mnras,
  423, 1805

\bibitem[{{Nordhaus} {et~al.}(2010){Nordhaus}, {Brandt}, {Burrows}, {Livne}, \&
  {Ott}}]{2010PhRvD..82j3016N}
{Nordhaus}, J., {Brandt}, T.~D., {Burrows}, A., {Livne}, E., \& {Ott}, C.~D.
  2010, \prd, 82, 103016

\bibitem[{{Noutsos} {et~al.}(2013){Noutsos}, {Schnitzeler}, {Keane}, {Kramer},
  \& {Johnston}}]{2013MNRAS.430.2281N}
{Noutsos}, A., {Schnitzeler}, D.~H.~F.~M., {Keane}, E.~F., {Kramer}, M., \&
  {Johnston}, S. 2013, \mnras, 430, 2281

\bibitem[{{Park} {et~al.}(2003){Park}, {Burrows}, {Garmire}, {Nousek},
  {Hughes}, \& {Williams}}]{2003ApJ...586..210P}
{Park}, S., {Burrows}, D.~N., {Garmire}, G.~P., {et~al.} 2003, \apj, 586, 210

\bibitem[{{Park} {et~al.}(2007){Park}, {Hughes}, {Slane}, {Burrows},
  {Gaensler}, \& {Ghavamian}}]{2007ApJ...670L.121P}
{Park}, S., {Hughes}, J.~P., {Slane}, P.~O., {et~al.} 2007, \apjl, 670, L121

\bibitem[{{Park} {et~al.}(2012){Park}, {Hughes}, {Slane}, {Burrows}, {Lee}, \&
  {Mori}}]{2012ApJ...748..117P}
---. 2012, \apj, 748, 117

\bibitem[{{Park} {et~al.}(2004){Park}, {Hughes}, {Slane}, {Burrows}, {Roming},
  {Nousek}, \& {Garmire}}]{2004ApJ...602L..33P}
---. 2004, \apjl, 602, L33

\bibitem[{{Paron} {et~al.}(2006){Paron}, {Reynoso}, {Purcell}, {Dubner}, \&
  {Green}}]{2006PASA...23...69P}
{Paron}, S.~A., {Reynoso}, E.~M., {Purcell}, C., {Dubner}, G.~M., \& {Green},
  A. 2006, \pasa, 23, 69

\bibitem[{{Pavan} {et~al.}(2014){Pavan}, {Bordas}, {P{\"u}hlhofer},
  {Filipovi{\'c}}, {De Horta}, {O'Brien}, {Balbo}, {Walter}, {Bozzo},
  {Ferrigno}, {Crawford}, \& {Stella}}]{2014A&A...562A.122P}
{Pavan}, L., {Bordas}, P., {P{\"u}hlhofer}, G., {et~al.} 2014, \aap, 562, A122

\bibitem[{{Pinheiro Gon{\c c}alves} {et~al.}(2011){Pinheiro Gon{\c c}alves},
  {Noriega-Crespo}, {Paladini}, {Martin}, \& {Carey}}]{2011AJ....142...47P}
{Pinheiro Gon{\c c}alves}, D., {Noriega-Crespo}, A., {Paladini}, R., {Martin},
  P.~G., \& {Carey}, S.~J. 2011, \aj, 142, 47

\bibitem[{{Posselt} {et~al.}(2017){Posselt}, {Pavlov}, {Slane}, {Romani},
  {Bucciantini}, {Bykov}, {Kargaltsev}, {Weisskopf}, \&
  {Ng}}]{2017ApJ...835...66P}
{Posselt}, B., {Pavlov}, G.~G., {Slane}, P.~O., {et~al.} 2017, \apj, 835, 66

\bibitem[{{Reed} {et~al.}(1995){Reed}, {Hester}, {Fabian}, \&
  {Winkler}}]{1995ApJ...440..706R}
{Reed}, J.~E., {Hester}, J.~J., {Fabian}, A.~C., \& {Winkler}, P.~F. 1995,
  \apj, 440, 706

\bibitem[{{Rest} {et~al.}(2011){Rest}, {Foley}, {Sinnott}, {Welch}, {Badenes},
  {Filippenko}, {Bergmann}, {Bhatti}, {Blondin}, {Challis}, {Damke}, {Finley},
  {Huber}, {Kasen}, {Kirshner}, {Matheson}, {Mazzali}, {Minniti}, {Nakajima},
  {Narayan}, {Olsen}, {Sauer}, {Smith}, \& {Suntzeff}}]{2011ApJ...732....3R}
{Rest}, A., {Foley}, R.~J., {Sinnott}, B., {et~al.} 2011, \apj, 732, 3

\bibitem[{{Reynoso} {et~al.}(2017){Reynoso}, {Cichowolski}, \&
  {Walsh}}]{2017MNRAS.464.3029R}
{Reynoso}, E.~M., {Cichowolski}, S., \& {Walsh}, A.~J. 2017, \mnras, 464, 3029

\bibitem[{{Reynoso} {et~al.}(1995){Reynoso}, {Dubner}, {Goss}, \&
  {Arnal}}]{1995AJ....110..318R}
{Reynoso}, E.~M., {Dubner}, G.~M., {Goss}, W.~M., \& {Arnal}, E.~M. 1995, \aj,
  110, 318

\bibitem[{{Russell} \& {Dopita}(1992)}]{1992ApJ...384..508R}
{Russell}, S.~C., \& {Dopita}, M.~A. 1992, \apj, 384, 508

\bibitem[{{Sagert} \& {Schaffner-Bielich}(2008)}]{2008A&A...489..281S}
{Sagert}, I., \& {Schaffner-Bielich}, J. 2008, \aap, 489, 281

\bibitem[{{Scheck} {et~al.}(2006){Scheck}, {Kifonidis}, {Janka}, \&
  {M{\"u}ller}}]{2006A&A...457..963S}
{Scheck}, L., {Kifonidis}, K., {Janka}, H.-T., \& {M{\"u}ller}, E. 2006, \aap,
  457, 963

\bibitem[{{Socrates} {et~al.}(2005){Socrates}, {Blaes}, {Hungerford}, \&
  {Fryer}}]{2005ApJ...632..531S}
{Socrates}, A., {Blaes}, O., {Hungerford}, A., \& {Fryer}, C.~L. 2005, \apj,
  632, 531

\bibitem[{{Tananbaum}(1999)}]{1999IAUC.7246....1T}
{Tananbaum}, H. 1999, \iaucirc, 7246

\bibitem[{{Tendulkar} {et~al.}(2012){Tendulkar}, {Cameron}, \&
  {Kulkarni}}]{2012ApJ...761...76T}
{Tendulkar}, S.~P., {Cameron}, P.~B., \& {Kulkarni}, S.~R. 2012, \apj, 761, 76

\bibitem[{{Tian} \& {Leahy}(2008)}]{2008MNRAS.391L..54T}
{Tian}, W.~W., \& {Leahy}, D.~A. 2008, \mnras, 391, L54

\bibitem[{{Tiengo} {et~al.}(2009){Tiengo}, {Esposito}, {Mereghetti}, {Israel},
  {Stella}, {Turolla}, {Zane}, {Rea}, {G{\"o}tz}, \&
  {Feroci}}]{2009MNRAS.399L..74T}
{Tiengo}, A., {Esposito}, P., {Mereghetti}, S., {et~al.} 2009, \mnras, 399, L74

\bibitem[{{Ugliano} {et~al.}(2012){Ugliano}, {Janka}, {Marek}, \&
  {Arcones}}]{2012ApJ...757...69U}
{Ugliano}, M., {Janka}, H.-T., {Marek}, A., \& {Arcones}, A. 2012, \apj, 757,
  69

\bibitem[{{Vasisht} \& {Gotthelf}(1997)}]{1997ApJ...486L.129V}
{Vasisht}, G., \& {Gotthelf}, E.~V. 1997, \apjl, 486, L129

\bibitem[{{Wachter} {et~al.}(2004){Wachter}, {Patel}, {Kouveliotou}, {Bouchet},
  {{\"O}zel}, {Tennant}, {Woods}, {Hurley}, {Becker}, \&
  {Slane}}]{2004ApJ...615..887W}
{Wachter}, S., {Patel}, S.~K., {Kouveliotou}, C., {et~al.} 2004, \apj, 615, 887

\bibitem[{{Wilms} {et~al.}(2000){Wilms}, {Allen}, \&
  {McCray}}]{2000ApJ...542..914W}
{Wilms}, J., {Allen}, A., \& {McCray}, R. 2000, \apj, 542, 914

\bibitem[{{Winkler} {et~al.}(1988){Winkler}, {Tuttle}, {Kirshner}, \&
  {Irwin}}]{1988srim.conf...65W}
{Winkler}, P.~F., {Tuttle}, J.~H., {Kirshner}, R.~P., \& {Irwin}, M.~J. 1988,
  in IAU Colloq. 101: Supernova Remnants and the Interstellar Medium, ed. R.~S.
  {Roger} \& T.~L. {Landecker}, 65

\bibitem[{{Winkler} {et~al.}(2009){Winkler}, {Twelker}, {Reith}, \&
  {Long}}]{2009ApJ...692.1489W}
{Winkler}, P.~F., {Twelker}, K., {Reith}, C.~N., \& {Long}, K.~S. 2009, \apj,
  692, 1489

\bibitem[{{Wongwathanarat} {et~al.}(2013){Wongwathanarat}, {Janka}, \&
  {M{\"u}ller}}]{2013A&A...552A.126W}
{Wongwathanarat}, A., {Janka}, H.-T., \& {M{\"u}ller}, E. 2013, \aap, 552, A126

\bibitem[{{Wongwathanarat} {et~al.}(2017){Wongwathanarat}, {Janka},
  {M{\"u}ller}, {Pllumbi}, \& {Wanajo}}]{2017ApJ...842...13W}
{Wongwathanarat}, A., {Janka}, H.-T., {M{\"u}ller}, E., {Pllumbi}, E., \&
  {Wanajo}, S. 2017, \apj, 842, 13

\bibitem[{{Woosley}(1987)}]{1987IAUS..125..255W}
{Woosley}, S.~E. 1987, in IAU Symposium, Vol. 125, The Origin and Evolution of
  Neutron Stars, ed. D.~J. {Helfand} \& J.-H. {Huang}, 255--270

\end{thebibliography}

%
%
%

\floattable
\begin{deluxetable}{ccccccc}
\tabletypesize{\tiny}
\tablecaption{Basic properties of the selected six SNRs\label{tab:targets}}
\tablewidth{0pt}
\tablehead{
\colhead{Parameter} & \colhead{Cassiopeia~A} & \colhead{G292.0+1.8} & 
\colhead{Puppis~A} & \colhead{Kes~73} & \colhead{RCW~103} & \colhead{N49} 
}
\startdata
Distance (kpc) & 3.4$^{+0.3}_{-0.1}$ [1] & 6.2$\pm$0.9 [5] & 2.2$^{+0.3}_{-0.9}$ [9,10] & 8.5$^{+1.3}_{-1.0}$ [15] & 3.3$^{+1.3}_{-0.2}$ [18] & 50 [23]\\
Age (yr) & 336$\pm$19 [2] & 2990$\pm$60 [6] & 4450$\pm$750 [11] & 750$\pm$250 [15] & 2000$^{+1050}_{-650}$ [19] & $\sim$4800 [24] \\
R$_{\rm FS}$ (pc) & 2.5 & 7.7 & 16 & 5.8 & 4.3 & 9.7 \\
R$_{\rm RS}$ (pc) & 1.6 [3] & 3.8 [7] & $\sim$9.6$^a$ & $\sim$4.1$^a$ & $\sim$2.8$^a$ & $\sim$5.5$^a$ \\
R$_{\rm RS}$ / Age (km\,s$^{-1}$) & $\sim$4700 & $\sim$1200 & $\sim$2100$^a$ & $\sim$5400 & $\sim$1400$^a$ & $\sim$1100$^a$ \\
Position of the NS & $\alpha=$23:23:27.9 [4] & $\alpha=$11:24:39.1 [8] & $\alpha=$8:21:57.3 [11] & $\alpha=$18:41:19.3 [16] & $\alpha=$16:17:36.3 [20] & $\alpha=$05:26:00.9 [25] \\
 & $\delta=$58:48:42.8 [4] & $\delta=$-59:16:20.5 [8] & $\delta=$-43:00:17.4 [11] & $\delta=$-8:45:56.0 [16] & $\delta=$-51:02:25.0 [20] & $\delta=$-66:04:36.3 [25] \\
Period (s) & N.A. & 0.135 [8] & 0.112 [12] & 11.8 [17] & 24000 [21] & 8.05 [26] \\
Period derivative (s\,s$^{-1}$) & N.A. & N.A. & 9.3$\times$10$^{-18}$ [13] & 4.1$\times$10$^{-11}$ [17] & $< 1.6\times10^{-9}$ [22] & 3.8$\times$10$^{-11}$ [26] \\
Center of expansion & $\alpha=$23:23:27.77$\pm$0.05 [2] & $\alpha=$11:24:34.4$\pm$0.5 [6] & $\alpha=$8:22:27.6$\pm$3.0 [14] & N.A. & N.A. & N.A. \\
 & $\delta=$58:48:49.4$\pm$0.4 [2] & $\delta=$-59:15:51$\pm$5 [6] & $\delta=$-42:57:28$\pm$60 [14] & N.A. & N.A. & N.A. \\
\enddata
\tablecomments{$^a$Since no direct measurement exists, we inferred the values from an evolutionary model by \citet{2003ApJ...593L..23C} with an assumption of the ejecta mass of 5\,M$_\odot$, a mass-loss rate of 2$\times$10$^{-5}$\,M\,yr$^{-1}$, and a stellar wind speed of 10\,km\,s$^{-1}$.  References are as follows: [1] \citet{1995ApJ...440..706R} [2] \citet{2006ApJ...645..283F} [3] \citet{2001ApJ...552L..39G} [4] \citet{1999IAUC.7246....1T} [5] \citet{2003ApJ...594..326G} [6] \citet{2009ApJ...692.1489W} [7] \citet{2015ApJ...800...65B} [8] \citet{2003ApJ...591L.139H} [9] \citet{1995AJ....110..318R} [10] \citet{2017MNRAS.464.3029R} [11] \citet{2012ApJ...755..141B} [12] \citet{2009ApJ...695L..35G} [13] \citet{2013ApJ...765...58G} [14] \citet{1988srim.conf...65W} [15] \citet{2008MNRAS.391L..54T} [16] \citet{2004ApJ...615..887W} [17] \citet{1997ApJ...486L.129V} [18] \citet{2006PASA...23...69P} [19] \citet{1997PASP..109..990C} [20] \citet{2000IAUC.7350....2G} [21] \citet{2006Sci...313..814D} [22] \citet{2011MNRAS.418..170E} [23] \citet{1999PASP..111..775F} [24] \citet{2012ApJ...748..117P} [25] \citet{2003ApJ...585..948K}. [26]\citet{2009MNRAS.399L..74T}.  }
\end{deluxetable}

\floattable
\begin{deluxetable}{lcc}
\tabletypesize{\tiny}
\tablecaption{Observations used in this paper\label{tab:data}}
\tablewidth{0pt}
\tablehead{
\colhead{Target} & \colhead{{\it Chandra}} & \colhead{{\it XMM-Newton}} 
}
\startdata
Cassiopeia~A & 114, 4638 & ... \\
G292.0+1.8 & 6677, 6679 & ... \\
Puppis~A & 12548, 13183 & 0113020101, 0150150101, 0150150201, 0150150301, 0303530101, 0606280101, 0606280201, 0690700201 \\
Kes~73 & 729, 16950, 17668, 17692, 17693 & ...\\
RCW~103 & 11823, 17460 & ... \\
N49 & 10123, 10806, 10807, 10808 & ... \\
\enddata
\tablecomments{The numbers indicate the observation IDs.}
\end{deluxetable}

\floattable
\begin{deluxetable}{lcccccccc}
\tabletypesize{\tiny}
\tablecaption{Best-fit parameters for representative spectra in Cas~A\label{tab:casa_param}}
\tablewidth{0pt}
\tablehead{
\colhead{Parameter} & \colhead{IME 1} & \colhead{Fe 1} & \colhead{O}& \colhead{CSM} & \colhead{Nonthermal}}
\startdata
{\tt TBabs} & Galactic absorption & Galactic absorption & Galactic absorption & Galactic absorption & Galactic absorption \\
$N_{\rm H}$ (10$^{22}$\,cm$^{-2}$) & 2.9$\pm$0.1 & 1.31$^{+0.03}_{-0.04}$ & 1.66$^{+0.05}_{-0.04}$ & 2.0$^{+0.2}_{-0.1}$ & 1.32$^{+0.03}_{-0.06}$ \\
\hline
{\tt vpshock} & O-rich ejecta &  & O-rich ejecta &  &  \\
$kT_{\rm e}$ (keV) & 3.2$^{+0.7}_{-0.4}$ & --- & 1.9$^{+0.1}_{-0.6}$ & --- & --- \\
C=N=O ($Z_\odot$) & 70000 ($>$30000) & --- & 61$^{+38}_{-3}$ & --- & --- \\
Ne ($Z_\odot$) & 0 ($<$210) & --- & 0 ($<$0.3) & --- & --- \\
Mg ($Z_\odot$) & 1100 ($>$800) & --- & 7.6$^{+17.9}_{-0.5}$ & --- & --- \\
log($n_{\rm e}$t/cm$^{-3}$\,s) & 9.74$^{+0.08}_{-0.11}$ & --- & 10.23$^{+0.05}_{-0.04}$ & --- & --- \\
Redshift (10$^{-3}$) & -17.0$^{+0.1}_{-0.2}$ & --- & -8.0$\pm$0.2 & --- & --- \\
Volume EM (10$^{55}$\,cm$^{-3}$) & 0.012$^{+5.3}_{-0.007}$ & --- & 2.8$\pm$0.2 & --- & --- \\
\hline
{\tt vpshock} & IME-rich ejecta & Fe-rich ejecta & IME-rich ejecta & CSM & CSM \\
$kT_{\rm e}$ (keV) & 1.7$\pm$0.1 & 2.0$^{+0.3}_{-0.2}$ & 2.3$^{+0.2}_{-0.1}$ & 3.3$^{+0.7}_{-0.8}$ & 5.1$\pm$1.5 \\
Ne ($Z_\odot$) & 1 (fixed) & 1 (fixed) & 1 (fixed) & 0.8$^{+0.4}_{-0.3}$ & 0.4$^{+0.3}_{-0.1}$ \\
Mg ($Z_\odot$) & 1 (fixed) & 1 (fixed) & 1 (fixed) & 0.7$\pm$0.1 & 0.2$\pm$0.1 \\
Si ($Z_\odot$) & 33$^{+11}_{-7}$ & 6.1 ($>$2.5) & 16.5$^{+0.3}_{-0.4}$ & 1.5$^{+0.5}_{-0.3}$ & 1.1$^{+0.7}_{-0.3}$ \\
S ($Z_\odot$) & 22$^{+7}_{-4}$ & 5.8 ($>$2.5) & 9.5$\pm$0.4 & 1.3$^{+0.6}_{-0.3}$ & 0.95$^{+0.61}_{-0.32}$ \\
Ar ($Z_\odot$) & 21$^{+8}_{-3}$ & 9.5$^{+102}_{-5}$ & 10.2$\pm$1.2 & 1 (fixed) & Linked to S \\
Ca ($Z_\odot$) & 24$^{+9}_{-7}$ & 10$^{+37}_{-7}$ & 11.2$\pm$3.4 & 1 (fixed) & Linked to S \\
Fe=Ni ($Z_\odot$) & 2.6$^{+1.0}_{-0.8}$ & 45$^{+570}_{-24}$ & 1 (fixed) & 1 (fixed) & 0.23$^{+0.21}_{-0.14}$ \\
log($n_{\rm e}$t/cm$^{-3}$\,s) & 11.07$^{+0.11}_{-0.09}$ & 11.46$^{+0.16}_{-0.12}$ & 10.60$\pm$0.02 & 10.59$^{+0.11}_{-0.05}$ & 10.93$^{+0.11}_{-0.09}$ \\
Redshift (10$^{-3}$) & Linked to O-rich comp. & -0.54$^{+0.68}_{-0.13}$ & Linked to O-rich comp. & -8.0$\pm$0.4 & -2.7$^{+0.1}_{-2.4}$ \\
Volume EM (10$^{55}$\,cm$^{-3}$) & 17$\pm$4 & 0.7$^{+0.8}_{-0.4}$ & 13.5$^{+0.3}_{-2.0}$ & 410$\pm$10 & 260$\pm$10 \\
\hline
{\tt power-law} & & & & &\\
$\Gamma$ & --- & 5.0$^{+0.6}_{-0.5}$ & --- & 2.7$^{+0.5}_{-0.2}$ & 2.2$\pm$0.1 \\
Norm (ph\,keV$^{-1}$\,cm$^{-2}$\,s$^{-1}$ at 1 keV) & --- & 0.00057$^{+0.00014}_{-0.00012}$ & --- & 0.0018$^{+0.0001}_{-0.0003}$ & 0.0011$\pm$0.0002 \\
\hline
$\chi^2$/d.o.f. & 427/319 & 198/132 & 367/254 & 372/343 & 399/375 \\
\enddata
\tablecomments{The region names in the first row correspond to those in Fig.~\ref{fig:specregs}.  The errors quoted represent 90\% statistical uncertainties.  Elemental abundances are relative to the solar values \citep{2000ApJ...542..914W}.  Other elemental abundances are fixed to the solar values.  The ionization parameters are fitted maxima with the lower limits being fixed at zero in the {\tt vpshock} model \citep{2001ApJ...548..820B}.  The volume EM is defined as $\int n_{\rm e} n_{\rm H} dV$, where $n_{\rm e}$ is the electron density, $n_{\rm H}$ is the hydrogen density, and $dV$ is the volume of the plasma. 
}
\end{deluxetable}

\floattable
\begin{deluxetable}{lcccccccc}
\tabletypesize{\tiny}
\tablecaption{Best-fit parameters for representative spectra in G292.0+1.8\label{tab:g292_param}}
\tablewidth{0pt}
\tablehead{
\colhead{Parameter} & \colhead{Si 1} & \colhead{O 1} & \colhead{CSM} & \colhead{PWN} }
\startdata
{\tt TBabs} & Galactic absorption & Galactic absorption & Galactic absorption & Galactic absorption \\
$N_{\rm H}$ (10$^{22}$\,cm$^{-2}$) & 0.65$\pm$0.07 & 0.90$^{+0.01}_{-0.04}$ & 0.53$\pm$0.01 & 0.49$^{+0.06}_{-0.03}$ \\
\hline
{\tt vpshock} & CSM & CSM & CSM & \\
$kT_{\rm e}$ (keV) & 0.08 ($<$0.15) & 0.08 ($<$0.22) & 0.08 ($<$0.22) & --- \\
Ne ($Z_\odot$) & 1 (fixed) & 1 (fixed) & 1.3$\pm$0.1 & --- \\
Mg ($Z_\odot$) & 1 (fixed) & 1 (fixed) & 1.1$\pm$0.1 & --- \\
Si ($Z_\odot$) & 1 (fixed) & 1 (fixed) & 0.7$\pm$0.1 & --- \\
S ($Z_\odot$) & 1 (fixed) & 1 (fixed) & 0.6$\pm$0.1 & --- \\
Fe=Ni ($Z_\odot$) & 1 (fixed) & 1 (fixed) & 0.50$\pm$0.03 & --- \\
log($n_{\rm e}$t/cm$^{-3}$\,s) & 11.3 (fixed) & 11.3 (fixed) & 11.42$^{+0.05}_{-0.07}$ & --- \\
Redshift (10$^{-3}$) & 0 (fixed) & 0 (fixed) & -3.6$\pm$0.4 & --- \\
Volume EM (10$^{58}$\,cm$^{-3}$) & 1.6$^{+1.7}_{-1.4}$ & 8.3$^{+2.6}_{-2.8}$ & 0.239$^{+0.007}_{-0.004}$ & --- \\
\hline
{\tt vpshock} & O-rich ejecta & O-rich ejecta & & O-rich ejecta \\
$kT_{\rm e}$ (keV) & 0.85$^{+0.09}_{-0.08}$ & 0.81$^{+0.10}_{-0.05}$ & --- & 0.52$^{+0.06}_{-0.02}$ \\
O ($Z_\odot$) & 133$^{+105}_{-53}$ & 73$^{+101}_{-3}$ & --- & 56$^{+42}_{-34}$ \\
Ne ($Z_\odot$) & 0 $<$82 & 86$^{+86}_{-2}$ & --- & 79$^{+186}_{-17}$ \\
Mg ($Z_\odot$) & 85$^{+3}_{-26}$ & 44$^{+2}_{-1}$ & --- & 52$^{+73}_{-13}$ \\
log($n_{\rm e}$t/cm$^{-3}$\,s) & 10.95$^{+0.02}_{-0.06}$ & 11.49$^{+0.06}_{-0.05}$ & --- & $>$12 \\
Redshift (10$^{-3}$) & -9.9$^{+0.4}_{-0.3}$ & -5.7$\pm$0.1 & --- & -9.8$\pm$0.3 \\
Volume EM (10$^{55}$\,cm$^{-3}$) & 1.1$^{+0.5}_{-0.4}$ & 6.5$^{+6.3}_{-3.3}$ & --- & 3.9$\pm$2.3 \\
\hline
{\tt vpshock} & IME-rich ejecta & IME-rich ejecta & & \\
$kT_{\rm e}$ (keV) & Linked to O-rich comp. & 0.36$^{+0.09}_{-0.05}$ & --- & --- \\
Si ($Z_\odot$) & 3.0$\pm$0.3 & 2700 ($>$1200) & --- & --- \\
S ($Z_\odot$) & 1.7$\pm$0.8 & Linked to Si & --- & --- \\
Ar ($Z_\odot$) & 0.26 ($<$1.3) & Linked to Si & --- & --- \\
Ca ($Z_\odot$) & Linked to Ar & Linked to Si & --- & --- \\
Fe ($Z_\odot$) & 0.4 ($<$1.2) & 1 (fixed) & --- & --- \\
log($n_{\rm e}$t/cm$^{-3}$\,s) & Linked to O-rich comp. & Linked to O-rich comp. & --- & --- \\
Redshift (10$^{-3}$) & Linked to O-rich comp. & Linked to O-rich comp. & --- & --- \\
Volume EM (10$^{55}$\,cm$^{-3}$) & 56$^{+4}_{-6}$ & 0.21$^{+0.19}_{-0.14}$ & --- & --- \\
\hline
{\tt power-law} & & & & \\
$\Gamma$ & --- & --- & 1.6$\pm$0.2 & 2.42$\pm$0.04 \\
Norm (ph\,keV$^{-1}$\,cm$^{-2}$\,s$^{-1}$ at 1 keV) & --- & --- & 0.00011$\pm$0.00005 & 0.00091$\pm$0.00005 \\
\hline
$\chi^2$/d.o.f. & 263/160 & 280/146 & 318/243 & 439/330 \\
\enddata
\tablecomments{Same notes as in Table~\ref{tab:casa_param}. 
}
\end{deluxetable}

\floattable
\begin{deluxetable}{lcccccccc}
\tabletypesize{\tiny}
\tablecaption{Best-fit parameters for representative spectra in Puppis~A\label{tab:pupa_param}}
\tablewidth{0pt}
\tablehead{
\colhead{Parameter} & \colhead{IME} & \colhead{O} & \colhead{CSM 1} & \colhead{CSM 2} }
\startdata
{\tt TBabs} & Galactic absorption & Galactic absorption & Galactic absorption & Galactic absorption \\
$N_{\rm H}$ (10$^{22}$\,cm$^{-2}$) & 0.33$\pm$0.01 & 0.24$\pm$0.06 & 0.17$\pm$0.01 & 0.22$^{+0.01}_{-0.02}$ \\
\hline
{\tt vpshock} & CSM & & CSM & CSM \\
$kT_{\rm e}$ (keV) & 0.20$^{+0.01}_{-0.02}$ & --- & 0.59$^{+0.02}_{-0.01}$ & 0.73$^{+0.05}_{-0.02}$ \\
O ($Z_\odot$) & 1 (fixed) & --- & 0.59$^{+0.02}_{-0.01}$ & 0.73$^{+0.05}_{-0.02}$ \\
Ne ($Z_\odot$) & 1 (fixed) & --- & 1.18$^{+0.02}_{-0.04}$ & 0.50$^{+0.04}_{-0.02}$ \\
Mg ($Z_\odot$) & 1 (fixed) & --- & 0.90$\pm$0.03 & 0.65$^{+0.04}_{-0.03}$ \\
Si ($Z_\odot$) & 1 (fixed) & --- & 0.87$\pm$0.06 & 0.74$\pm$0.05 \\
S ($Z_\odot$) & 1 (fixed) & --- & 0.7$\pm$0.2 & 0.73$\pm$0.15 \\
Fe=Ni ($Z_\odot$) & 1 (fixed) & --- & 0.76$\pm$0.02 & 0.44$^{+0.05}_{-0.01}$ \\
log($n_{\rm e}$t/cm$^{-3}$\,s) & 13 ($>$12) & --- & 11.11$^{+0.01}_{-0.03}$ & 11.11$\pm$0.02  \\
Redshift (10$^{-3}$) & 0 (fixed) & --- & 1.9$\pm$0.1 & 1.9$\pm$0.1 \\
Volume EM (10$^{58}$\,cm$^{-3}$) & 0.92$^{+1.0}_{-0.2}$ & --- & 200$\pm$1 & 103$^{+2}_{-15}$ \\
\hline
{\tt vpshock} & O-rich ejecta & O-rich ejecta &  & \\
$kT_{\rm e}$ (keV) & 0.83$^{+0.01}_{-0.04}$ & 0.77$^{+0.18}_{-0.13}$ & --- & --- \\
C=N=O ($Z_\odot$) & 1.15$^{+0.11}_{-0.02}$ & 9.7$^{+3.9}_{-3.0}$ & --- & --- \\
Ne ($Z_\odot$) & 2.07$^{+0.08}_{-0.04}$ & 10.3$^{+3.8}_{-3.0}$ & --- & --- \\
Mg ($Z_\odot$) & 1.86$\pm$0.04 & 5.9$^{+3.5}_{-2.7}$ & --- & --- \\
log($n_{\rm e}$t/cm$^{-3}$\,s) & 11.02$\pm$0.01 & 10.51$^{+0.13}_{-0.12}$ & --- & --- \\
Redshift (10$^{-3}$) & -2.0$\pm$0.1 & -0.54$^{+1.9}_{-0.1}$ & --- & --- \\
Volume EM (10$^{55}$\,cm$^{-3}$) & 340$\pm$10 & 2.1$^{+0.9}_{-0.6}$ & --- & --- \\
\hline
{\tt vpshock} & IME-rich ejecta & & & \\
$kT_{\rm e}$ (keV) & 0.60$^{+0.4}_{-0.3}$ & --- & --- & --- \\
Si ($Z_\odot$) & 526$^{+500}_{-25}$ & --- & --- & --- \\
S ($Z_\odot$) & 410$^{+356}_{-52}$ & --- & --- & --- \\
Ar ($Z_\odot$) & 0 ($<$105) & --- & --- & --- \\
Ca ($Z_\odot$) & Linked to Ar & --- & --- & --- \\
log($n_{\rm e}$t/cm$^{-3}$\,s) & 11.32$^{+0.12}_{-0.10}$ & --- & --- & --- \\
Redshift (10$^{-3}$) & Linked to O-rich comp. & --- & --- & --- \\
Volume EM (10$^{55}$\,cm$^{-3}$) & 3.2$^{+1.3}_{-0.2}$ & --- & --- & --- \\
\hline
$\chi^2$/d.o.f. & 228/144 & 127/95 & 246/113 & 217/133 \\
\enddata
\tablecomments{Same notes as in Table~\ref{tab:casa_param}. 
}
\end{deluxetable}

\floattable
\begin{deluxetable}{lc}
\tabletypesize{\tiny}
\tablecaption{Best-fit parameters for the spectrum from the whole Kes~73\label{tab:kes73_param}}
\tablewidth{0pt}
\tablehead{
\colhead{Parameter} & \colhead{Entire SNR}}
\startdata
{\tt TBabs} & Galactic absorption \\
$N_{\rm H}$ (10$^{22}$\,cm$^{-2}$) & 3.4$^{+0.11}_{-0.06}$ \\
\hline
{\tt vpshock} & CSM \\
$kT_{\rm e}$ (keV) & 1.78$^{+0.17}_{-0.11}$  \\
log($n_{\rm e}$t/cm$^{-3}$\,s) & 11.28$^{+0.06}_{-0.09}$  \\
Redshift (10$^{-3}$) & 2.7$\pm$0.1 \\
Volume EM (10$^{58}$\,cm$^{-3}$) & 2.5$^{+1.3}_{-0.4}$ \\
\hline
{\tt vpshock} & IME-rich ejecta \\
$kT_{\rm e}$ (keV) & 0.65$\pm$0.08  \\
C=N=O ($Z_\odot$) & 3.0$^{+2.0}_{-1.4}$ \\
Ne ($Z_\odot$) & 0.4$^{+0.3}_{-0.2}$ \\
Mg ($Z_\odot$) & 1.3$^{+0.4}_{-0.1}$ \\
Si ($Z_\odot$) & 2.2$^{+1.1}_{-0.5}$ \\
S ($Z_\odot$) & 2.2$^{+1.4}_{-0.5}$  \\
Ar=Ca ($Z_\odot$) & 2.6$\pm$0.5 \\
Fe ($Z_\odot$) & 0.5$^{+0.2}_{-0.1}$ \\
log($n_{\rm e}$t/cm$^{-3}$\,s) & 11.52$^{+0.21}_{-0.04}$ \\
Redshift (10$^{-3}$) & Linked to CSM comp. \\
Volume EM (10$^{58}$\,cm$^{-3}$) & 17.4$^{+5.5}_{-2.1}$ \\
\hline
$\chi^2$/d.o.f. & 443/330 \\
\enddata
\tablecomments{Same notes as in Table~\ref{tab:casa_param}. 
}
\end{deluxetable}

\floattable
\begin{deluxetable}{lcc}
\tabletypesize{\tiny}
\tablecaption{Best-fit parameters for the representative spectra of RCW~103\label{tab:rcw103_param}}
\tablewidth{0pt}
\tablehead{
\colhead{Parameter} & \colhead{Ejecta} & \colhead{CSM1}}
\startdata
{\tt TBabs} & Galactic absorption & Galactic absorption \\
$N_{\rm H}$ (10$^{22}$\,cm$^{-2}$ & 1.20$^{+0.06}_{-0.03}$ & 1.37$\pm$0.09 \\
\hline
{\tt vpshock} & CSM & CSM\\
$kT_{\rm e}$ (keV) & 0.50$^{+0.03}_{-0.05}$ & 0.47$\pm$0.03 \\
C=N=O ($Z_\odot$) & 1 (fixed) & 1.3$^{+3.8}_{-0.9}$ \\
Ne ($Z_\odot$) & 1 (fixed) & 1.0$^{+2.8}_{-0.5}$ \\
Mg ($Z_\odot$) & 1 (fixed) & 1.0$^{+1.6}_{-0.2}$ \\
log($n_{\rm e}$t/cm$^{-3}$\,s) & 11.3 (fixed) & 11.27$^{+0.16}_{-0.21}$  \\
Redshift (10$^{-3}$) & 0 (fixed) & 6.0$^{+0.7}_{-0.4}$ \\
Volume EM (10$^{55}$\,cm$^{-3}$) & 26$^{+2}_{-4}$ & 99$^{+17}_{-43}$ \\
\hline
{\tt vpshock} & IME-rich ejecta \\
$kT_{\rm e}$ (keV) & 0.96$^{+0.09}_{-0.07}$ & --- \\
Mg ($Z_\odot$) & 0 ($<$3.3) & --- \\
Si ($Z_\odot$) & 29$^{+93}_{-7}$ & --- \\
S ($Z_\odot$) & 18$^{+2}_{-5}$ & --- \\
Ar=Ca ($Z_\odot$) & 6$^{+45}_{-3}$ & --- \\
Fe ($Z_\odot$) & 18$^{+111}_{-6}$ & --- \\
log($n_{\rm e}$t/cm$^{-3}$\,s) & 11.96$^{+0.31}_{-0.20}$ & --- \\
Redshift (10$^{-3}$) & 11$^{+2}_{-1}$ & --- \\
Volume EM (10$^{55}$\,cm$^{-3}$) & 0.2$^{+0.6}_{-0.1}$ & ---  \\
\hline
$\chi^2$/d.o.f. & 200/143 & 101/98\\
\enddata
\tablecomments{Same notes as in Table~\ref{tab:casa_param}. 
}
\end{deluxetable}

\floattable
\begin{deluxetable}{lcc}
\tabletypesize{\tiny}
\tablecaption{Best-fit parameters for the representative spectra of N49\label{tab:n49_param}}
\tablewidth{0pt}
\tablehead{
\colhead{Parameter} & \colhead{IME 2} & \colhead{CSM 1}}
\startdata
{\tt vphabs} & LMC absorption & LMC absorption \\
$N_{\rm H}$ (10$^{22}$\,cm$^{-2}$) & 0.18$\pm$.09 & 0 ($<$0.05) \\
\hline
{\tt TBabs} & Galactic absorption & Galactic absorption \\
$N_{\rm H}$ in our Galactic (10$^{22}$\,cm$^{-2}$) & 0.06 (fixed) & 0.06 (fixed) \\
\hline
{\tt vpshock} & CSM \\
$kT_{\rm e}$ (keV) & 1.45$^{+0.39}_{-0.22}$ & 0.65$^{+0.08}_{-0.60}$ \\
O ($Z_\odot$) & 0.7 (fixed) & 0.44$^{+0.40}_{-0.20}$\\
Ne ($Z_\odot$) & 0.9 (fixed) & 0.57$^{+0.32}_{-0.18}$ \\
Mg ($Z_\odot$) & 0.7 (fixed) & 0.47$^{+0.28}_{-0.17}$ \\
log($n_{\rm e}$t/cm$^{-3}$\,s) & 11.88$^{+0.73}_{-0.39}$ & 11.75$\pm$0.21 \\
Redshift (10$^{-3}$) & 0 (fixed) & 0 (fixed) \\
Volume EM (10$^{58}$\,cm$^{-3}$) & 0.38$\pm$0.05 & 0.57$^{+0.18}_{-0.16}$\\
\hline
{\tt vpshock} & IME-rich ejecta \\
$kT_{\rm e}$ (keV) & 0.68$^{+0.06}_{-0.04}$ & --- \\
Mg ($Z_\odot$) & 3.6$^{+354}_{-2.4}\times$10$^4$ & --- \\
Si ($Z_\odot$) & 10.5 ($<$3.8) & --- \\
S=Ar=Ca ($Z_\odot$) & 10.5 ($<$3.7)   & --- \\
Fe ($Z_\odot$) & 1.0$^{+102}_{-0.6}\times$10$^4$ & --- \\
log($n_{\rm e}$t/cm$^{-3}$\,s) & 13 ($<$12) & --- \\
Redshift (10$^{-3}$) & 0 (fixed) & --- \\
Volume EM (10$^{55}$\,cm$^{-3}$) & 32$^{+166}_{-19}$ & --- \\
\hline
$\chi^2$/d.o.f. & 101/94 & 63/66 \\
\enddata
\tablecomments{Same notes as in Table~\ref{tab:casa_param}.  The absorption in our Galaxy is fixed at the value in the literature \citep{2012ApJ...748..117P}. The elemental abundances in the absorption model for the LMC are fixed at the values in the literature \citep{1992ApJ...384..508R,1998ApJ...505..732H}. For the Spectrum 1, the abundances of the CSM component are fixed at typical values measured in some CSM-dominated regions.
}
\end{deluxetable}

\floattable
\begin{deluxetable}{ccccccc}
\tabletypesize{\tiny}
\tablecaption{Summary of our X-ray measurements\label{tab:results}}
\tablewidth{0pt}
\tablehead{
\colhead{Parameter} & \colhead{Cassiopeia~A} & \colhead{G292.0+1.8} & 
\colhead{Puppis~A} & \colhead{Kes~73} & \colhead{RCW~103} & \colhead{N49} 
}
\startdata
Center of the IME mass & $\alpha=$23:23:28.1$\pm$0.6 & $\alpha=$11:24:30.0$\pm$0.7 & $\alpha=$8:22:57.9$\pm$1.0 & $\alpha=$18:41:19.7$\pm$0.2 & $\alpha=$16:17:.34.8$\pm$1.1 & $\alpha=$5:26:00.57$\pm$0.16 \\
 & $\delta=$58:49:03.4$\pm$5.0 & $\delta=$-59:15:40.0$\pm$4.0 & $\delta=$-42:51:09$\pm$12 & $\delta=$-4:56:19.9$\pm$2.5 & $\delta=$-51:02:51.7$\pm$7.9 & $\delta=$-66:05:01.9$\pm$1.0 \\
Center of the X-ray boundary & $\alpha=$23:23:27.9$\pm$0.3 & $\alpha=$11:24:33.1$\pm$0.2 & $\alpha=$8:22:15.7$\pm$4.5 & $\alpha=$18:41:19.6$\pm$0.1 & $\alpha=$16:17:36.3$\pm$0.8 & $\alpha=$5:25:59.57$\pm$0.16 \\
 & $\delta=$58:48:56.2$\pm$3.5 & $\delta=$-59:15:51.1$\pm$2.5 & $\delta=$-43:02:00$\pm$10 & $\delta=$-4:56:17.6$\pm$1.8 & $\delta=$-51:02:36.7$\pm$7.2 & $\delta=$-66:04:56.4$\pm$1.0 \\
Opening angle between NS and CoM(IME) & 159$^{+21}_{-18}$ & 159$\pm$6 & 158$\pm$1 & 169$^{+62}_{-48}$ & 134$^{+30}_{-20}$ & 111$\pm$5 \\
IME mass$^a$ (M$_\odot$) & $\sim$6.4$\times$10$^{-2}$ ($\sim$0.14) & $\sim$2.9$\times$10$^{-2}$ ($\sim$0.67) & $\sim$0.2 ($\sim$0.3) & $\sim$1.1$\times$10$^{-2}$ ($\sim$0.6) & $\sim$1.2$\times$10$^{-2}$ ($\sim$4$\times$10$^{-2}$) & $\sim$0.36 ($\sim$2) \\
Asymmetry parameter ($\alpha_{\rm IME}$) & 0.18$\pm$0.04 & 0.36$\pm$0.05 & 0.84$\pm$0.02 & 0.14$\pm$0.02 & 0.25$\pm$0.02 & 0.52$\pm$0.04 \\
Plane-of-the-sky NS speed (km\,s$^{-1}$) & 340$^{+90}_{-70}$ [1] & 450$\pm$70 [2] & 900$^{+200}_{-400}$ [3,4] & 400$\pm$150 & 100$^{+60}_{-30}$ & 1100$\pm$50 \\
\enddata
\tablecomments{The uncertainties indicate one-sigma confidence levels.  $^a$The masses are calculated based on the assumption that the plasma depth is 10\% of the SNR radius employing a SNR evolutionary model \citep{2003ApJ...593L..23C}. These masses strongly depend on the assumed plasma depth and/or the filling factor so that the values are subject to large uncertainties of a factor of a few.  Moreover, the metal masses strongly depend on contributions of hydrogen and helium that are difficult to measure with X-ray spectra, if the plasma is rich in metals.  Therefore, we give masses expected for pure metal plasmas in parenthes as upper limits of IME masses.  Uncertainties on NS kick speeds for Kes~73, RCW~103, and N49 should be considered to be lower limits, given that we assume that the CoX is the explosion site, which is subject to systematic uncertainties.  References are as follows: [1] \citet{2006ApJ...645..283F} [2] \citet{2009ApJ...692.1489W} [3] \citet{1988srim.conf...65W} [4] \citet{2012ApJ...755..141B}.}
\end{deluxetable}

\clearpage

\end{document}